\RequirePackage{fixltx2e}
\documentclass[notitlepage,twocolumn,prd,floatfix,showpacs,aps, a4paper]{revtex4-1}
\usepackage[nodayofweek]{datetime}
\usepackage{paralist}
\usepackage{amsmath}
\usepackage{mathtools}
\usepackage{natbib}
\usepackage{color}
\usepackage{nicefrac}
\usepackage{hyperref}
\usepackage{graphicx}
\usepackage[caption=false]{subfig}
\newcommand{\kms}{\,\textrm{km s}^{-1}}

\usepackage{multirow}
\usepackage{tabulary}
\usepackage{array}
\usepackage{comment}

\begin{document}

\title{Parametrizing the local dark matter speed distribution: a detailed analysis}

\author{Bradley J. Kavanagh}
\email{Electronic address: ppxbk2@nottingham.ac.uk}
\affiliation{School of Physics \& Astronomy,University of Nottingham, University Park, Nottingham, NG7 2RD, UK}

\date{\today}

\begin{abstract}
In a recent paper, a new parametrization for the dark matter (DM) speed distribution $f(v)$ was proposed for use in the analysis of data from direct detection experiments. This parametrization involves expressing the \textit{logarithm} of the speed distribution as a polynomial in the speed $v$. We present here a more detailed analysis of the properties of this parametrization. We show that the method leads to statistically unbiased mass reconstructions and exact coverage of credible intervals. The method performs well over a wide range of DM masses, even when finite energy resolution and backgrounds are taken into account. We also show how to select the appropriate number of basis functions for the parametrization. Finally, we look at how the speed distribution itself can be reconstructed, and how the method can be used to determine if the data are consistent with some test distribution. In summary, we show that this parametrization performs consistently well over a wide range of input parameters and over large numbers of statistical ensembles and can therefore reliably be used to reconstruct both the DM mass and speed distribution from direct detection data.
\end{abstract}


\maketitle

\section{Introduction}

The dark matter (DM) paradigm has enjoyed much success in explaining a wide range of astronomical observations (for a review, see e.g.\ Ref.~\cite{Bertone:2005}). As yet no conclusive evidence has been provided for the identity of particle DM, though there are a variety of candidates, including the supersymmetric neutralino \cite{Jungman:1996}, sterile neutrinos \cite{Dodelson:1994}, axions \cite{Duffy:2009} and the lightest Kaluza-Klein particle \cite{Kolb:1984}. Here, we focus on the search for particles which belong to the generic class of Weakly Interacting Massive Particles (WIMPs). Direct detection experiments \cite{Goodman:1985, Drukier:1986} aim to measure the energies of nuclear recoils induced by WIMP DM in the Galactic halo. Under standard assumptions about the DM halo, this data can be used to extract the WIMP mass and interaction cross section, allowing us to check for consistency with other search channels (such as indirect detection \cite{Lavalle:2012} and collider experiments \cite{Battaglia:2010}) and to probe underlying models of DM.

Direct detection experiments are traditionally analyzed within the framework of the Standard Halo Model (SHM), in which WIMPs are assumed to have a Maxwell-Boltzmann speed distribution in the Galactic frame. The impact of uncertainties in the WIMP speed distribution has been much studied (see e.g.\ Refs.~\cite{Green:2010, Peter:2011, Fairbairn:2012}), leading to the conclusion that such uncertainties may introduce a bias into any reconstruction of the WIMP mass from direct detection data. As yet, the speed distribution is unknown, while a number of proposals have been put forward for its form, including analytic parametrizations (e.g. Ref.~\cite{Lisanti:2010}) and distributions reconstructed from the potential of the Milky Way \cite{Bhattacharjee:2012} or from N-body simulations \cite{Vogelsberger:2009, Kuhlen:2010,Kuhlen:2012, Mao:2012}. Recent results from N-body simulations which attempt to include the effects of baryons on structure formation also report the possible presence of a dark disk in the Milky Way \cite{Read:2009, Read:2010, Kuhlen:2013}. With such a wide range of possibilities, we should take an agnostic approach to the speed distribution, not only to avoid introducing bias into the analysis of data, but also with the hope of measuring the speed distribution and thereby probing the formation history of the Milky Way.

Several methods of evading these uncertainties have been proposed. These include simultaneously fitting the parameters of the SHM and dark matter properties \cite{Strigari:2009,Peter:2009}, fitting to empirical forms of the speed distribution (e.g.\ Ref.~\cite{Pato:2011}) and fitting to a self-consistent distribution function \cite{Pato:2013}. However, these methods typically require that the speed distribution can be well fitted by a particular functional form. More model-independent methods, such as fitting the moments of the speed distribution \cite{Drees:2007, Drees:2008} or using a step-function speed distribution \cite{Peter:2011}, have also been presented. However, these methods can still introduce a bias into the measurement of the WIMP mass and perform less well with the inclusion of realistic experimental energy thresholds.

In a recent paper \cite{Kavanagh:2013a} (hereafter referred to as Paper 1), a new parametrization of the speed distribution was presented, which allowed the WIMP mass to be extracted from hypothetical direct detection data without prior knowledge of the speed distribution itself. Paper 1 demonstrated this for a WIMP of mass 50 GeV, using several underlying distribution functions. In the present paper, we extend this analysis to a wider range of masses. We also aim to demonstrate the statistical properties of the method and show how realistic experimental parameters affect its performance. Finally, we will also elaborate on some of the technical details of the method and assess its ability to reconstruct the underlying WIMP speed distribution.

Section~\ref{sec:DDRate} of this paper explains the direct detection event rate formalism and presents the parametrization of the speed distribution introduced in Paper 1. In Sec.~\ref{sec:ParameterRecon}, the methodology for testing the parametrization is outlined. In Section \ref{sec:Parametrization}, we consider the choice and number of basis functions for the method. We then study the performance of the method as a function of input WIMP mass (Sec.~\ref{sec:mass}) and when Poisson fluctations in the data are taken into account (Sec.~\ref{sec:stats}). In Sec.~\ref{sec:Recon}, we demonstrate how the speed distribution can be extracted from this parametrization and examine whether or not different distribution functions can be distinguished. Finally, we summarize in Sec.~\ref{sec:Conclusions} the main results of this paper.

\section{Direct detection event rate}
\label{sec:DDRate}

Dark matter direct detection experiments aim to measure the energies $E$ of nuclear recoils induced by interactions with WIMPs in the Galactic halo. Calculation of the event rate at such detectors has been much studied (e.g. Refs.~\cite{Goodman:1985, Drukier:1986, Lewin:1996,Jungman:1996}). For a target nucleus with nucleon number $A$, interacting with a WIMP of mass $m_\chi$, the event rate per unit detector mass is given by:
\begin{equation}
\label{eq:Rate}
\frac{\textrm{d}R}{\textrm{d}E} = \frac{\rho_0 \sigma_p}{2 m_\chi \mu_{\chi p}^2} A^2 F^2(E) \eta(v_\textrm{min})\,,
\end{equation}
where $\rho_0$ is the local dark matter mass density, $\sigma_p$ is the WIMP-proton spin-independent cross section and the reduced mass is defined as $\mu_{A B} = m_A m_B/(m_A + m_B)$. The Helm form factor $F^2(E)$ \cite{Helm:1956} describes the loss of coherence of spin-independent scattering due to the finite size of the nucleus. A wide range of possible interactions have been considered in the literature, including inelastic \cite{Smith:2001}, isospin-violating \cite{Kurylov:2003} and more general non-relativistic interactions \cite{Fan:2010, Fitzpatrick:2012, Fitzpatrick:2013}. We focus here on the impact of the WIMP speed distribution on the direct detection event rate. We therefore restrict ourselves to considering only spin-independent scattering, which is expected to dominate over the spin-dependent contribution for heavy nuclei, due to the $A^2$ enhancement in the rate.

Information about the WIMP velocity distribution $f(\textbf{v})$ is encoded in the function $\eta$, sometimes referred to as the mean inverse speed,
\begin{equation}
\label{eq:eta}
\eta(v_\textrm{min}) = \int_{v > v_\textrm{min}} \frac{f(\textbf{v})}{v} \, \textrm{d}^3\textbf{v}\,,
\end{equation}
where $\textbf{v}$ is the WIMP velocity in the reference frame of the detector. The integration is performed only over those WIMPs with sufficient speed to induce a nuclear recoil of energy $E$. The minimum required speed for a target nucleus of mass $m_N$ is
\begin{equation}
\label{eq:v_min}
v_\textrm{min}(E) = \sqrt{\frac{m_N E}{2\mu_{\chi N}^2}}\,.
\end{equation}

We distinguish between the directionally averaged velocity distribution
\begin{equation}
f(v) = \oint f(\textbf{v}) \, \textrm{d}\Omega_{\textbf{v}}\,,
\end{equation}
and the 1-dimensional speed distribution
\begin{equation}
f_1(v) = \oint f(\textbf{v}) v^2 \textrm{d}\Omega_{\textbf{v}}\,.
\end{equation}
The distribution function should in principle be time-dependent, due to the motion of the Earth around the Sun. However, this is expected to be a percent-level effect (for a review, see e.g. Ref.~\cite{Freese:2013}) and we therefore assume that $f_1(v)$ is time independent in the present work.

\begin{figure}[t]
  \includegraphics[width=0.49\textwidth]{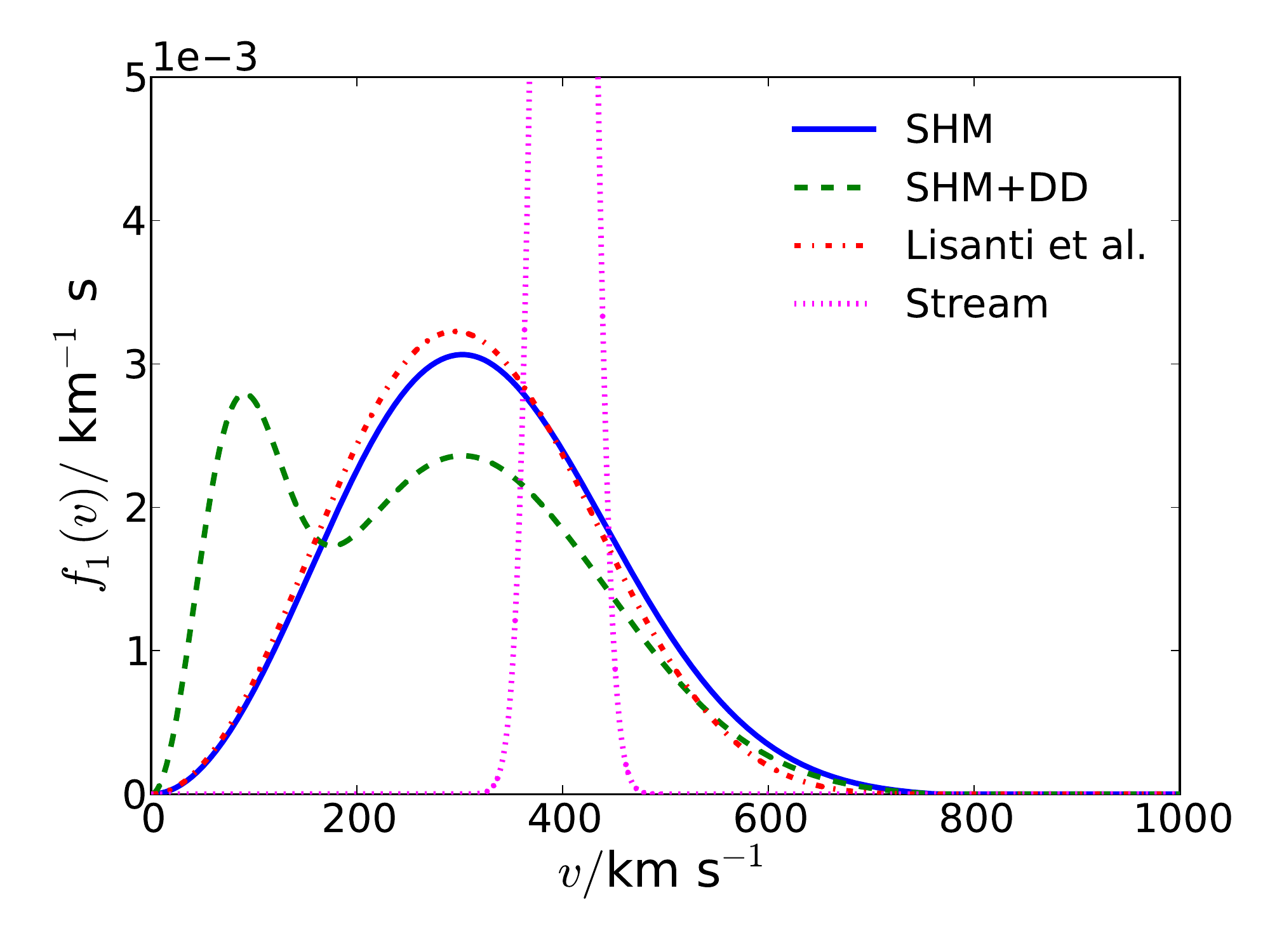}
  \caption{Several of the benchmark speed distributions used in this work. They are defined in Eqs.~\ref{eq:gaussian} and \ref{eq:lisanti} with parameters from Tab.~\ref{tab:distributions}. These distributions are the SHM (solid blue), SHM+DD (dashed green), Lisanti et al. (dot-dashed red) and the stream (dotted magenta).}
  \label{fig:Ensemble_distributions}
\end{figure}

We consider several benchmark speed distributions in this work, including the SHM and the SHM with the addition of a moderate dark disk which accounts for 23\% of the total WIMP density \cite{Kuhlen:2013}. We model the speed distributions as combinations of Gaussian functions in the Earth frame
\begin{equation}
\label{eq:gaussian}
g(\textbf{v}) = N \exp\left(-\frac{(\textbf{v} - \textbf{v}_\textrm{lag})^2}{2\sigma_v^2}\right) \Theta(v_\textrm{esc} - |\textbf{v} - \textbf{v}_\textrm{lag}|)\,,
\end{equation}
where $\textbf{v}_\textrm{lag}$ specifies the peak velocity of the distribution in the Earth frame and $\sigma_v$ the velocity dispersion. We truncate the distribution above the escape speed $v_\textrm{esc}$ in the Galactic frame and the factor $N$ is required to satisfy the normalization condition (Eq.~\ref{eq:normalization}). We use the value $v_\textrm{esc} = 544 \kms$, which lies within the 90\% confidence limits obtained from the RAVE survey \cite{RAVE:2007, RAVE:2013}. In addition, we also use the speed distribution of Lisanti et al.~\cite{Lisanti:2010}, which has the following form in the Earth's frame:
\begin{equation}
\label{eq:lisanti}
f(\textbf{v}) = N \left[\exp\left(\frac{v_\textrm{esc}^2 - |\textbf{v} - \textbf{v}_0|^2}{k v_0^2}\right) -1\right]^k \Theta(v_\textrm{esc} - |\textbf{v} - \textbf{v}_0|)\,.
\end{equation}
We use the parameter values $k = 2$ and $v_0 = 220 \kms$ in this work. We summarize in Tab.~\ref{tab:distributions} the different speed distributions considered. We also plot several of these in Fig.~\ref{fig:Ensemble_distributions} for reference.

\begin{table}[t]
  \setlength{\extrarowheight}{2pt}
  \setlength{\tabcolsep}{3pt}
  \begin{center}
	\begin{tabular}{m{3cm}|ccc}
	Speed distribution benchmark & Fraction & $v_\textrm{lag} / \kms$ & $\sigma_v / \kms$ \\
	\hline\hline
	SHM & 1 & 220 & 156 \\
	\hline
	\multirow{2}{*}{SHM+DD} & 0.77 & 220 & 156 \\
	& 0.23 & 50 & 50 \\
	\hline
	Stream & 1 & 400 & 20 \\
	\hline
	\multirow{2}{*}{Bump} & 0.97 & 220 & 156 \\
	& 0.03 & 500 & 20 \\
	\hline
	\multirow{2}{*}{Double-peak} & 0.5 & 200 & 20 \\
	& 0.5 & 400 & 20 \\
	\hline\hline
	Lisanti et al. & & $v_0 = 220 \kms$ & $k = 2$
	\end{tabular}
  \end{center}
\caption{Summary of speed distribution benchmarks used in this work. Some benchmarks are modelled as mixtures of two gaussian components (defined in Eq.~\ref{eq:gaussian}), for which we give the fractional contribution of each component (labelled `Fraction'). The remaining parameters are defined in Eqs.~\ref{eq:gaussian} and \ref{eq:lisanti} and the accompanying text. The `bump' and `double-peak' distributions are discussed in Sec.~\ref{sec:Parametrization}.}
\label{tab:distributions}
\end{table}

In Paper 1, a parametrization for the WIMP speed distribution was introduced, for use in the analysis of direct detection data. The parametrization of Paper 1 has the form:
\begin{equation}
\label{eq:parametrization}
f_1(v) = v^2 \exp\left\{ -\sum_{k =0}^{N-1} a_k P_k(v)\right\}\,
\end{equation}
where $P_k(v)$ is some basis of polynomial functions of v. We fit the coefficients $\left\{a_1, ..., a_{N-1}\right\}$ using data, and fix $a_o$ by normalization
\begin{equation}
\label{eq:normalization}
a_0 = \ln\left(\int_{0}^\infty v^2 \exp\left\{ -\sum_{k = 1}^{N-1} a_k P_k(v)\right\} \, \textrm{d}v\right)\,.
\end{equation}
This form of parametrization ensures that the distribution function $f_1(v)$ is everywhere positive and can be used to fit an arbitrary underlying directionally-averaged distribution function (given a sufficiently large number of polynomial basis functions). We explore in Sec.~\ref{sec:Parametrization} which basis functions should be used in the parametrization, as well as how many basis functions are required.

\section{Parameter Reconstruction}
\label{sec:ParameterRecon}
In order to assess the performance of the parametrization method, we attempt to reconstruct the WIMP mass $m_\chi$ and polynomial coefficients $\left\{a_1, ..., a_{N-1} \right\}$ using the nested sampling software \textsc{MultiNest} \cite{MultiNest1, MultiNest2, MultiNest3}. We also include the WIMP-proton spin-independent cross section $\sigma_p$ as a free parameter. However, we are forced to treat the cross section as a nuisance parameter. As has previously been noted \cite{Kavanagh:2012, Kavanagh:2013a}, taking an agnostic approach to the DM speed distribution means that we do not know what fraction of WIMPs lie above the energy thresholds of the experiments. While this does not adversely impact the reconstruction of the WIMP mass, it does result in a strong degeneracy, such that only lower limits can be placed on the cross section using such methods. In any case, the cross section appears in the event rate (Eq.~\ref{eq:Rate}) in the degenerate combination $\rho_0 \sigma_p$. Uncertainties on the local DM density $\rho_0$ are at present on the order of a factor of 2 (see e.g.~\cite{Iocco:2011, Bovy:2012, Zhang:2013, Nesti:2013}) and thus any reconstruction of the cross section would be subject to the same systematic uncertainty. In this work, we focus instead on reconstructing the WIMP mass and the shape of the speed distribution. For concreteness, we use the values $\sigma_p = 10^{-45} \textrm{ cm}^2$ and $\rho_0 = 0.3 \textrm{ GeV cm}^{-3}$ throughout this work.

\subsection{Experimental benchmarks}
\label{sec:experiments}

In order to generate mock data sets, we consider three idealized mock experiments, loosely based on detectors which are currently in development. As previous work has shown \cite{Kavanagh:2012, Peter:2013a}, the WIMP mass and speed distribution are degenerate when data from only a single experiment is considered. However, this degeneracy can be broken by including data from additional experiments with different nuclear target masses. The three target materials we consider here are Xenon, Argon and Germanium. We describe each experiment in terms of its nucleon number $A$, fiducial detector mass $m_\textrm{det}$, efficiency $\epsilon$ and energy sensitivity window $\left[E_\textrm{min}, E_\textrm{max}\right]$.
We incorporate the effects of detector sensitivity, analysis cuts and detector down-time into the value of the efficiency $\epsilon$, which we take to be energy independent for simplicity. We consider a total exposure time for all experiments of $t_\textrm{exp} = \textrm{ 2 years}$. The experimental parameter values used in this work are summarized in Tab.~\ref{tab:experiments}.

\begin{table}[t]
  \setlength{\extrarowheight}{2pt}
  \begin{center}
	\begin{tabular}{c|m{1.2cm}m{1.7cm}m{1.5cm}m{1.7cm}}
	Experiment  & Target Mass, $A$ & Detector Mass (fid.), $m_\textrm{det}$/kg & Efficiency, $\epsilon$ & Energy Range/keV\\
	\hline\hline
	Xenon  & 131  & 1100 \cite{Aprile:2012a} & 0.7 \cite{Aprile:2012b} & 7-45 \cite{Aprile:2010} \\
	Argon  & 40  & 1000 & 0.9 \cite{Benetti:2007} & 30-100 \cite{Grandi:2005} \\
        Germanium  & 73  & 150 \cite{Bauer:2013b} & 0.6 \cite{Bauer:2013a} & 8-100 \cite{Bauer:2013a} \\
	\end{tabular}
  \end{center}
\caption{Summary of experimental parameters used in this work, defined in Sec.~\ref{sec:experiments}. An exposure of $t_\textrm{exp} = 2 \textrm{ years}$ is used for all 3 experiments.}
\label{tab:experiments}
\end{table}

The exact parameter values we used in this work do not strongly impact the results we present. However, it is important to note that the total mass and exposure of the experiments will affect the total number of events observed. This in turn will affect the precision of the reconstructions. For example, we have chosen a total Argon mass of 1000 kg. This is the stated target for Argon-based experiments which are in development (e.g. Ref.~\cite{Badertscher:2013}), though at present typical fiducial masses for Argon prototypes are of the order of 100 kg \cite{Grandi:2005}. The data we have generated does not represent the `high-statistics' regime: across all three experiments the total number of events observed is roughly 200-300 with as few as 10 events in the Germanium detector for some scenarios. Using a smaller exposure (or equivalently a smaller interaction cross section) will reduce the precision of the results, but should not introduce any additional bias. We also briefly consider the impact of a \textit{larger} number of events in Sec.~\ref{sec:Recon}.

\subsection{Parameter sampling}

We make parameter inferences using a combination of Bayesian and frequentist statistics. Bayes theorem for the probability of a particular set of theoretical parameters $\boldsymbol{\Theta}$ given the observed data $\textbf{D}$ is:

\begin{equation}
P(\boldsymbol{\Theta}|\textbf{D}) = \frac{P(\boldsymbol{\Theta}) P(\textbf{D}|\boldsymbol{\Theta})}{P(\textbf{D})}\,,
\end{equation}
where $P(\boldsymbol{\Theta})$ is the prior on the parameters and $P(\textbf{D})$ is the Bayesian evidence, which acts as a normalizing factor and has no impact on parameter inference. We summarize the priors used in this work in Tab.~\ref{tab:priors}. We also summarize in Tab.~\ref{tab:MultiNest} the MultiNest sampling parameters used.

\begin{table}[t]
  \setlength{\extrarowheight}{2pt}
  \setlength{\tabcolsep}{3pt}
  \begin{center}
	\begin{tabular}{m{1in}|cc}
	Parameter & Prior type & Prior range\\
	\hline\hline
	$m_\chi / \textrm{ GeV}$ &  log-flat & $\left[10^{0}, 10^{3}\right]$\\
	$\sigma_p / \textrm{ cm}^2$ & log-flat & $\left[10^{-46}, 10^{-42}\right]$ \\
	$\left\{a_k\right\}$ & linear-flat & $\left[-50, 50\right]$ \\
        $R_{BG} / \textrm{dru}$ & log-flat & $\left[10^{-12}, 10^{-5}\right]$ \\
	\end{tabular}
  \end{center}
\caption{Summary of the priors on the parameters used in this work. The background rate $R_{BG}$ is defined in Sec.~\ref{sec:mass}, while the remaining parameters are defined in Sec.~\ref{sec:DDRate}.}
\label{tab:priors}
\end{table}

\begin{table}[t]
  \setlength{\extrarowheight}{2pt}
  \setlength{\tabcolsep}{3pt}
  \begin{center}
	\begin{tabular}{c|c}
        Parameter & Value \\
        \hline\hline
	$N_\textrm{live}$ & 10000 \\
	efficiency & 0.25 \\
	tolerance & $10^{-4}$ \\
	\end{tabular}
  \end{center}
\caption{Summary of the MultiNest sampling parameters used in this work.}
\label{tab:MultiNest}
\end{table}

The factor $P(\textbf{D}|\boldsymbol{\Theta})$ is simply the likelihood of the data given the parameters $\boldsymbol{\Theta}$. In Sec.~\ref{sec:Parametrization} and Sec.~\ref{sec:mass}, we consider the effects of varying the form of the parametrization and of varying the input WIMP mass. In order to eliminate the effects of Poisson noise, we use Asimov data \cite{Cowan:2013} for these sections. This means that we divide the energy window of each experiment into bins of width 1 keV. We then set the observed number of events $N_{o,i}$ in bin $i$ equal to the expected number of events $N_{e,i}$. In this case, we use the binned likelihood, calculated for $N_b$ energy bins:
\begin{equation}
\label{eq:binnedL}
\mathcal{L}_b = \prod_{i = 1}^{N_b} \frac{N_{e,i}^{N_{o,i}} \textrm{e}^{-N_{e,i}}}{N_{o,i}!}\,.
\end{equation}

In Sec.~\ref{sec:stats} and Sec.~\ref{sec:Recon}, we consider many realisations of data, including the effects of Poisson noise. We therefore use the extended likelihood which has previously been used by both the Xenon \cite{Aprile:2011} and CDMS \cite{Ahmed:2009} collaborations, which for a single experiment is given by:
\begin{equation}
\label{eq:unbinnedL}
\mathcal{L} = \frac{N_e^{N_o} \textrm{e}^{-N_e}}{N_o!} \prod_{i = 1}^{N_o} P(E_i)\,,
\end{equation}
where the expected number of events is given by:
\begin{equation}
\label{eq:N_expected}
N_e = \epsilon m_\textrm{det}t_\textrm{exp}\int_{E_\textrm{min}}^{E_\textrm{max}} \frac{\textrm{d}R}{\textrm{d}E}\, \textrm{d}E\,,
\end{equation}
and the normalised recoil spectrum is given by:
\begin{equation}
\label{eq:eventdistribution}
P(E) = \frac{ \epsilon m_\textrm{det} t_\textrm{exp}}{N_e} \frac{\textrm{d}R}{\textrm{d}E}\,.
\end{equation}
The total likelihood is then the product over all experiments considered.

Using nested sampling, we can extract the full posterior probability distribution of the parameters $P(\boldsymbol{\Theta}|\textbf{D})$, as well as the likelihood $\mathcal{L}(\Theta)$. However, we often want to make inferences not jointly for all parameters but for only a subset (treating the remaining as nuisance parameters). If we conceptually partition the parameter space into the parameters of interest $\boldsymbol{\psi}$ and the remaining nuisance parameters $\boldsymbol{\phi}$, we would like to make inferences about the values of $\boldsymbol{\psi}$, without reference to the values of $\boldsymbol{\phi}$. One option for doing this is to calculate the marginalized posterior distribution, obtained by integrating the posterior probability over the parameters we are not interested in:

\begin{equation}
P_m(\boldsymbol{\psi}) = \int P(\boldsymbol{\psi}, \boldsymbol{\phi}) \, \textrm{d}\boldsymbol{\phi}\,.
\end{equation}
This method performs well for small numbers of observations (compared to the number of free parameters in the fit). We take the mode of the distribution to be the reconstructed parameter value and construct p\% \textit{minimal} credible intervals, which include those parameter values with $P_m(\boldsymbol{\psi}) \geq h$, where $h$ is chosen such that p\% of the probability distribution lies within the interval. The marginalized posterior method is used in Sec.~\ref{sec:stats} and Sec.~\ref{sec:Recon}, where in some cases the number of events observed in an experiment is less than 10.

An alternative method is to calculate the profile likelihood. This is obtained by maximizing the full likelihood function over the nuisance parameters:
\begin{equation}
\label{eq:profilelikelihood}
\mathcal{L}_p(\boldsymbol{\psi}) = \max_{\boldsymbol{\phi}} \mathcal{L}(\boldsymbol{\psi},\boldsymbol{\phi})\,.
\end{equation}
For a large number of observations, we can take the value which maximizes $\mathcal{L}_p$ as the reconstructed value and construct confidence intervals using the asymptotic properties of the profile likelihood. We use the profile likelihood for parameter inferences in Sec.~\ref{sec:Parametrization} and Sec.~\ref{sec:mass}, as the Asimov data sets provide a large number of measurements of $N_{e,i}$ over a large number of bins. The profile likelihood can also lead to less noisy reconstructions than the marginalized posterior, especially when the dimensionality of the parameter space becomes high, as in Sec.~\ref{sec:Parametrization} and Sec.~\ref{sec:mass}.

\section{Testing the parametrization}
\label{sec:Parametrization}

\begin{figure}[t]
  \includegraphics[width=0.49\textwidth]{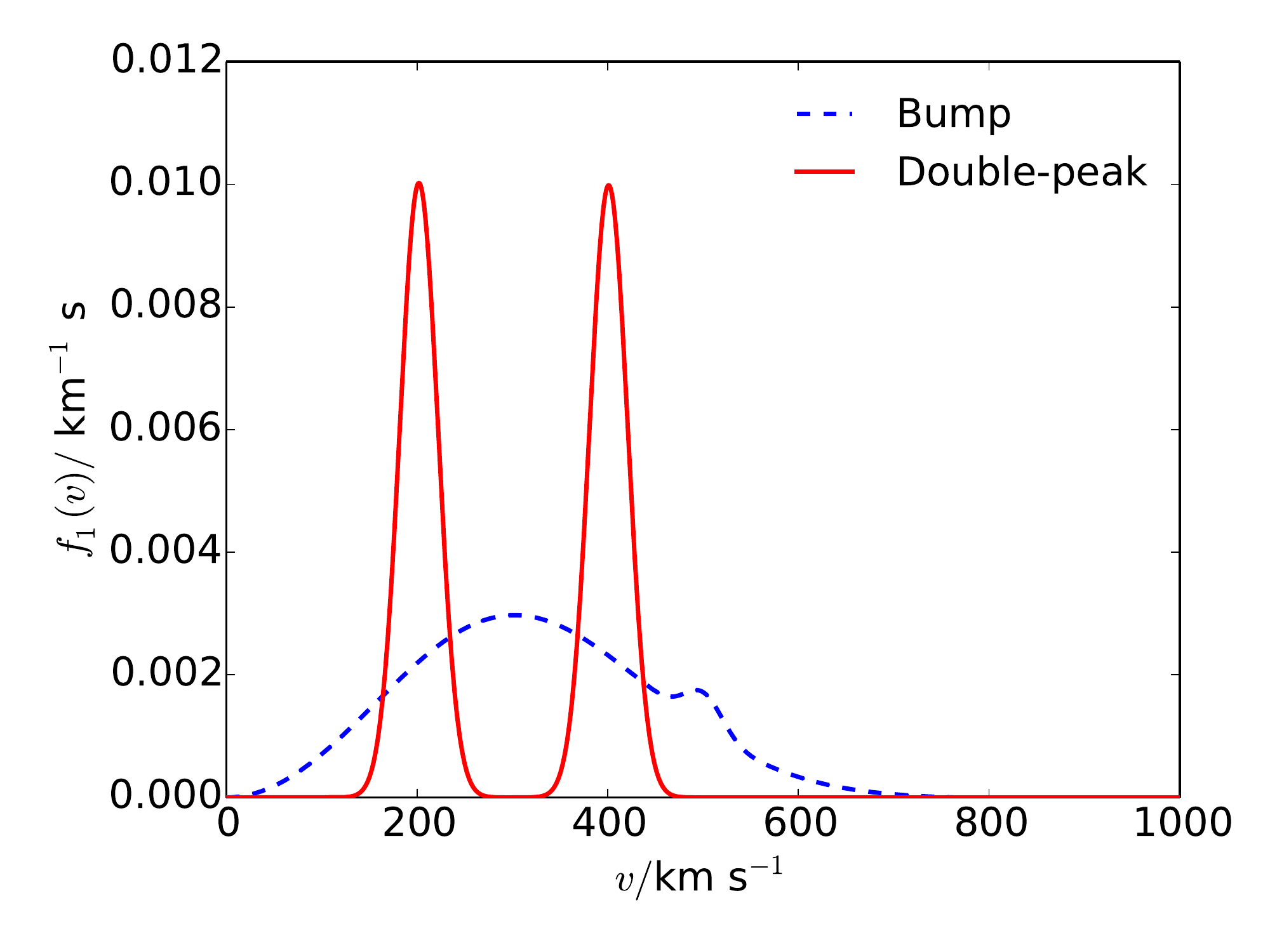}
  \caption{Benchmark speed distributions used in Sec.~\ref{sec:Parametrization} to test the performance of the parametrization as a function of the number and type of basis functions.}
  \label{fig:VaryingN_distributions}
\end{figure}

We now consider the two questions: \begin{inparaenum} \item how many basis functions are required and \item which polynomial basis should be used? \end{inparaenum} In order to answer these questions, we use the two benchmark distribution functions illustrated in Fig.~\ref{fig:VaryingN_distributions}. We have chosen these benchmarks not because they are necessarily realistic distribution functions but because they should be difficult to fit using standard techniques and fitting functions (e.g.~\cite{Lisanti:2010}). The first distribution (referred to as `bump') is a SHM distribution with the addition of a small bump, which contributes just 3\% of the total WIMP population and could correspond to a small sub-halo or stream \cite{Vogelsberger:2009}. This should be difficult to fit because it represents only a very small deviation from the standard scenario. The second distribution (referred to as `double-peak') has a sharp and rapidly varying structure, which we anticipate should be difficult to capture using a small number of basis functions.

\subsection{Varying the number of basis functions}

We first investigate how the reconstructed WIMP mass $m_\textrm{rec}$ and uncertainty varies with the number of basis functions $N$. For now, we fix our choice of basis to shifted Legendre polynomials, as used in Paper 1:

\begin{equation}
P_k(v) = L_k\left(2\frac{v}{v_\textrm{max}} - 1\right)\,,
\end{equation}
where $L_k$ is the Legendre polynomial of order $k$, and $v_\textrm{max}$ is a cut off for the parametrization. We should choose $v_\textrm{max}$ to ensure that $f_1(v)$ is negligible above the cut off. However, too high a choice of $v_\textrm{max}$ will result in $f_1(v)$ being close to zero over a large range of the parametrization, making fitting more difficult. We use the value $v_\textrm{max} = 1000 \kms$, while lies significantly above the Galactic escape speed.

\begin{figure}[t]
  \includegraphics[width=0.49\textwidth]{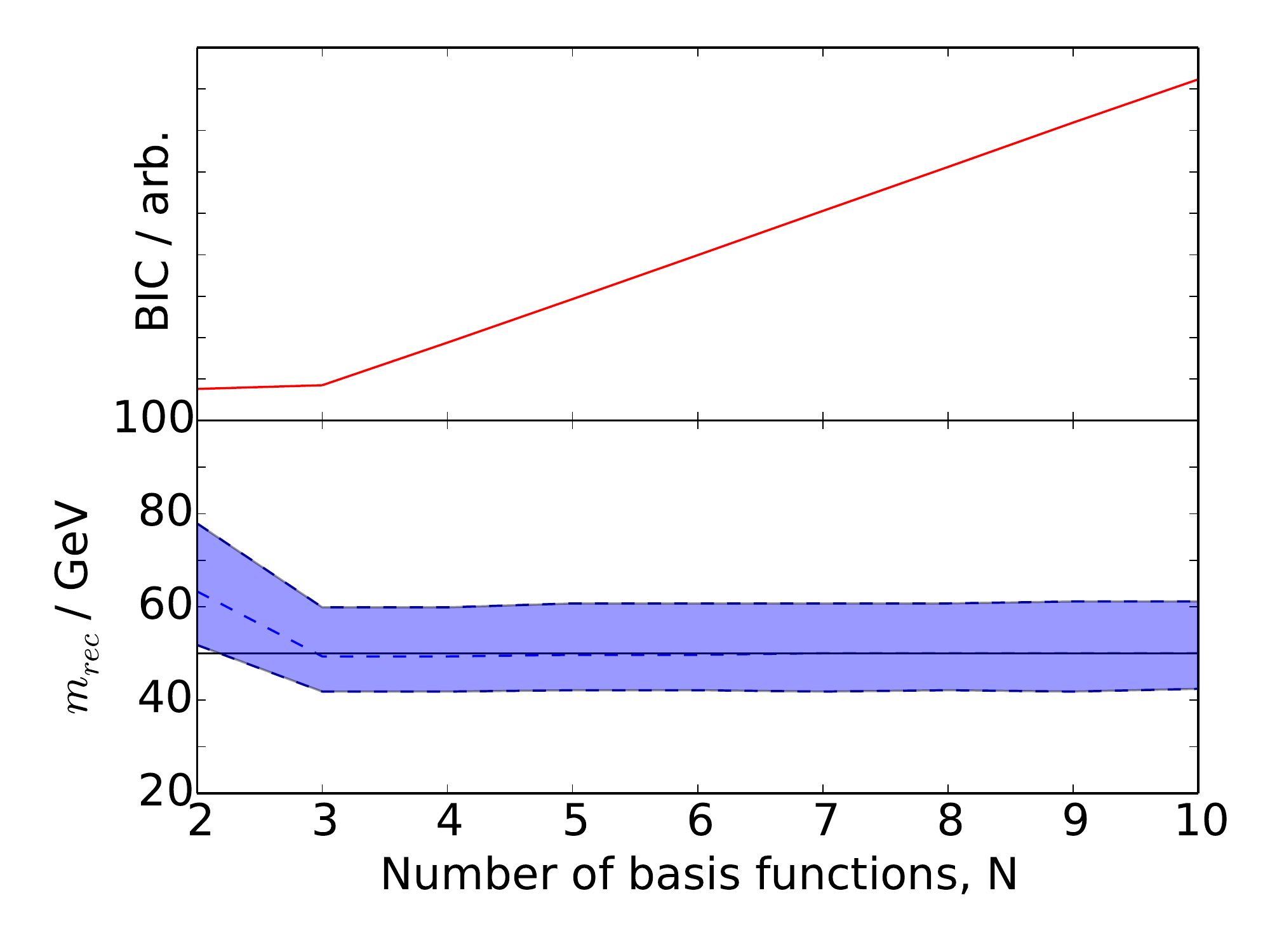}
  \caption{Bayesian information criterion (BIC) as a function of the number of basis functions for an underlying `bump' distribution function, 50 GeV WIMP and using Legendre polynomial basis functions (upper panel). Also shown (lower panel) are the reconstructed WIMP mass (dashed blue line), 68\% confidence interval (shaded blue region) and underlying WIMP mass (solid horizontal black line).}
  \label{fig:BUMP_LEG}
\end{figure}

The lower panel of Fig.~\ref{fig:BUMP_LEG} shows the best fit mass and 68\% confidence intervals as a function of $N$, using as input a WIMP of mass 50 GeV and the `bump' distribution function. The reconstructed mass very rapidly settles close to the true value, using as few as three basis functions. This is because adding the bump near $v \sim 500 \kms$ still leaves the mean inverse speed relatively smooth, so a large number of basis function are not required. The correct mass is reconstructed and we emphasize in the lower panel of Fig.~\ref{fig:BUMP_LEG} that the reconstruction is stable with the addition of more basis functions.

We should also consider how the quality of the fit changes as a function of $N$. We would expect that adding fit parameters should always lead to a better fit. Eventually, the fit should be good enough that adding additional basis functions will no longer improve it significantly. We can then be confident that our reconstruction is accurate and not an artifact of using too few basis functions. In order to investigate this, we utilise the Bayesian Information Criterion (BIC) \cite{Schwarz:1978}, which is given by:

\begin{equation}
BIC = 2N_p\textrm{ln}(N_m) - \textrm{ln}(\mathcal{L}_\textrm{max}) \, ,
\end{equation}
where $N_p$ is the number of free parameters, $N_m$ is the number of measurements or observations and $\mathcal{L}_\textrm{max}$ is the maximum likelihood value obtained in the reconstruction. For the case of binned data, $N_m$ corresponds simply to the total number of energy bins across all experiments. This criterion penalises the inclusion of additional free parameters and in comparing several models, we should prefer the one which minimises the BIC.

The upper panel of Fig.~\ref{fig:BUMP_LEG} shows the BIC (in arbitrary units) as a function of the number of basis functions for the `bump' distribution function. The BIC is comparable for the cases of $N=2$ and $N=3$, indicating that the quality of the fit is improved slightly by the addition of another basis function. However, adding further basis functions does not have a significant impact on the maximum likelihood, leading to an increase in the BIC. This coincides with the stabilization of the reconstructed mass around the true value and we conclude that only two or three basis functions are required to provide a good fit to the data.

\begin{figure}[t]
  \includegraphics[width=0.49\textwidth]{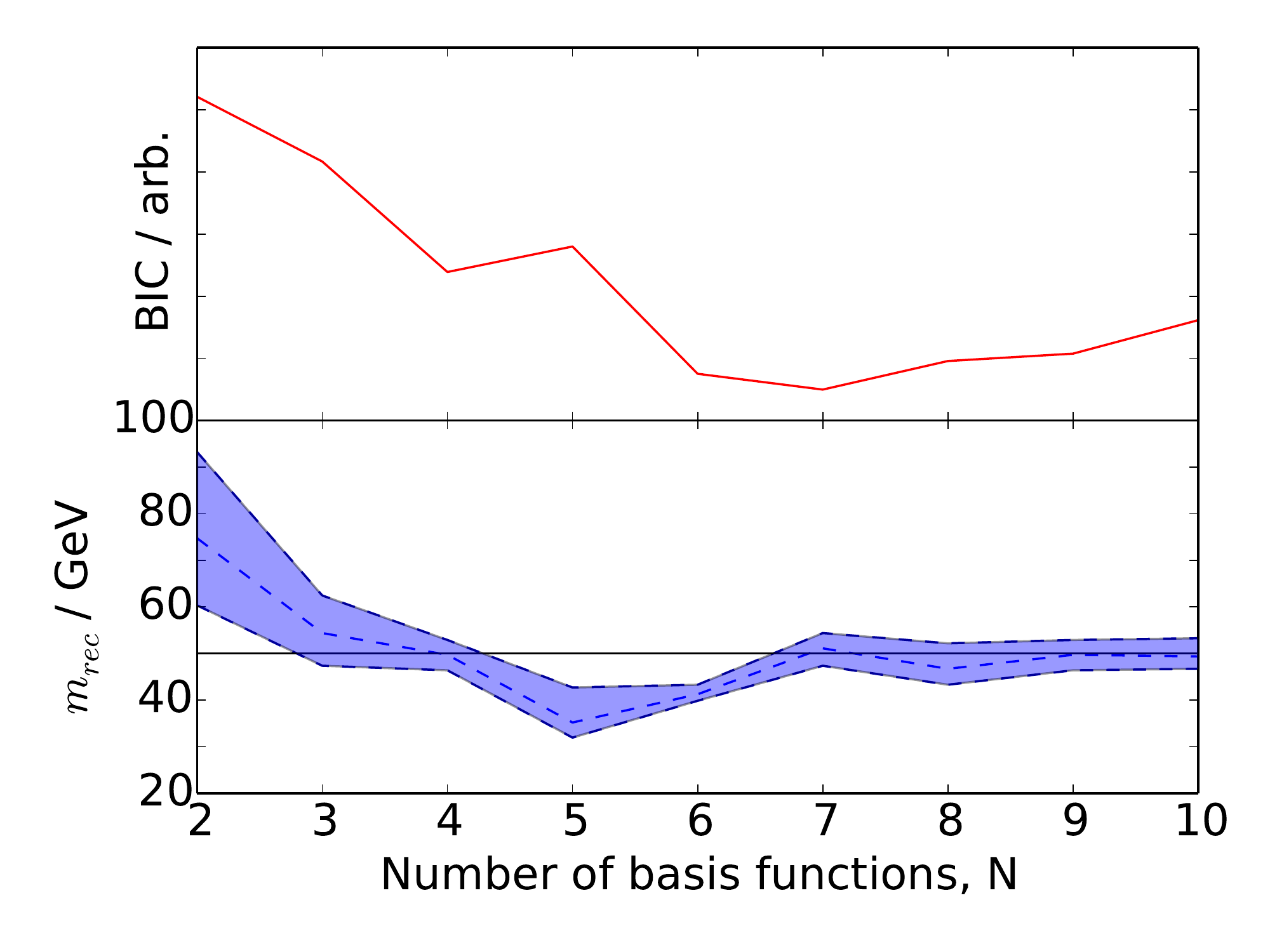}
  \caption{As Fig.~\ref{fig:BUMP_LEG} but for an underlying `double-peak' distribution function.}
  \label{fig:DP_LEG}
\end{figure}

Figure \ref{fig:DP_LEG} shows the corresponding results for the `double-peak' distribution function. Here, we note that the bias induced by using too small a number of basis functions is larger than for the case of the `bump' distribution, due to the more complicated structure in this case. The BIC is minimized for $N=7$, indicating that additional basis functions do not significantly improve the quality of the fit to data. This suggests that the shape of the speed distribution can be well fit by $N\geq7$ basis functions. As shown in the lower panel of Fig.~\ref{fig:DP_LEG}, the reconstruction of the WIMP mass is stable around the true mass for these values of $N$.

We propose that such a procedure should be used in the case of real data should a dark matter signal be observed at multiple detectors. We have shown that by analyzing the reconstructed mass as a function of $N$ we can recover the true mass and that by using the BIC we can be confident that we have obtained an adequate fit to data.

\subsection{Choice of basis functions}

We now consider the second question posed at the start of Sec.~\ref{sec:Parametrization}: which polynomial basis should be used? We see immediately that a naive power series of the form

\begin{equation}
\textrm{ln}f(v) \approx a_0 + a_1 v + a_2 v^2 + a_3 v^3 + ...\,,
\end{equation}
is not practical for the purposes of parameter estimation. Higher powers of $v$ will have rapidly growing contributions to $\textrm{ln} f$, meaning that the associated coefficients must be rapidly decreasing in order to suppress these contributions. Fitting to the SHM using just 5 terms, the range of values for the $a_k$ in the case of a simple power series would span around 13 orders of magnitude. Ideally, we would like to specify an identical prior on each of the coefficients. However, in this scenario this would result in a highly inefficient exploration of the parameter space when some of the terms are so small.

This problem can be significantly improved by rescaling $v$. We choose to rescale by a factor of $v_\textrm{max} = 1000 \kms$, and cut off the distribution function at $v_\textrm{max}$. The basis functions $(v/v_\textrm{max})^k$ are now less than unity by construction and the coefficients $a_k$ are now dimensionless:

\begin{equation}
\textrm{ln}f(v) \approx a_0 + a_1 (v/v_\textrm{max}) + a_2 (v/v_\textrm{max})^2 + a_3 (v/v_\textrm{max})^3 + ...\,.
\end{equation}

We now address the problem of \textit{conditioning} of the polynomial basis (see e.g.\ Refs.~\cite{Gautschi:1978, Wilkinson:1984}). Conditioning is a measure of how much the value of a polynomial changes, given a small change in the coefficients. For a well-conditioned polynomial, small changes in the coefficient are expected to lead to small changes in the value of the polynomial. This is ideal for parameter estimation as it leads to a more efficient exploration of the parameter space. Orthogonal polynomial basis functions typically have improved conditioning \cite{Gautschi:1978} and we consider two specific choices: the Legendre polynomials which have already been considered and the Chebyshev polynomials. The Chebyshev polynomials are used extensively in polynomial approximation theory \cite{Mason:2002} and are expected to be well conditioned \cite{Gautschi:1978}.

\begin{figure}[t]
  \includegraphics[width=0.49\textwidth]{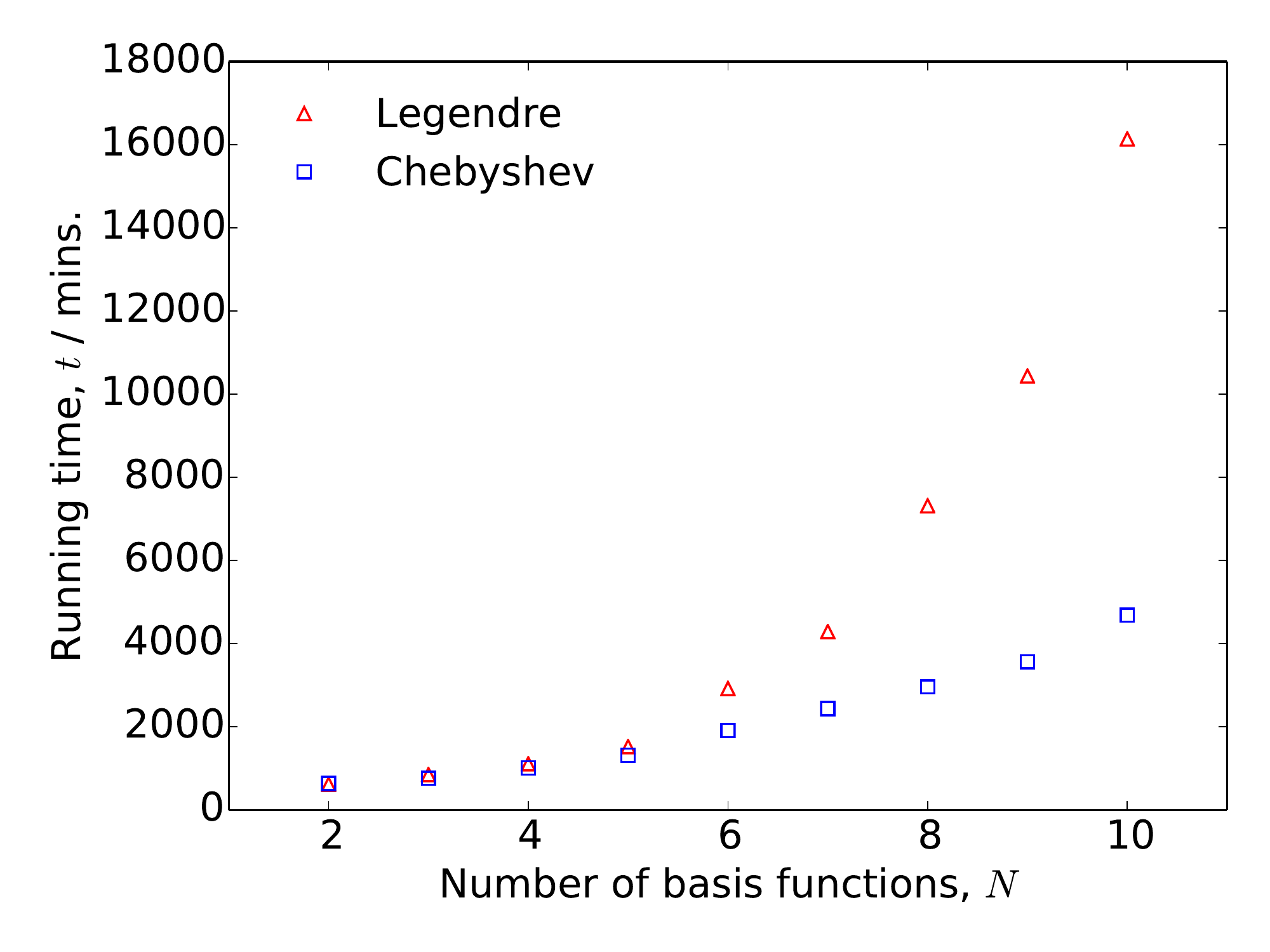}
  \caption{Time taken (using 4 processors in parallel) for the reconstruction of the `bump' benchmark, as a function of number of basis functions. The time taken using the Chebyshev basis (blue squares) grows more slowly with $N$ than for the Legendre basis (red triangles).}
  \label{fig:times}
\end{figure}

We have checked that the reconstruction results using Chebyshev polynomials are largely indistinguishable from the case of Legendre polynomials for both the `bump' and `double-peak' distributions and as a function of $N$. This leads us to conclude that the accuracy of the reconstruction is independent of the specific choice of basis. However, the reconstruction was much faster in the case of the Chebyshev basis. This is illustrated in Fig.~\ref{fig:times}, which shows the time taken for reconstruction of the `bump' benchmark as a function of $N$. The time taken grows much more slowly for the Chebyshev basis (roughly as $N^2$) than for the Legendre basis (roughly as $N^3$). We have also checked that this difference is not an artifact of how we calculate the basis functions. These results indicate that this choice of basis provides both reliable and efficient reconstruction for the WIMP mass and we therefore use the Chebyshev basis in the remainder of this work.

\section{Varying $m_\chi$}
\label{sec:mass}

\begin{figure}[t]
  \includegraphics[width=0.49\textwidth]{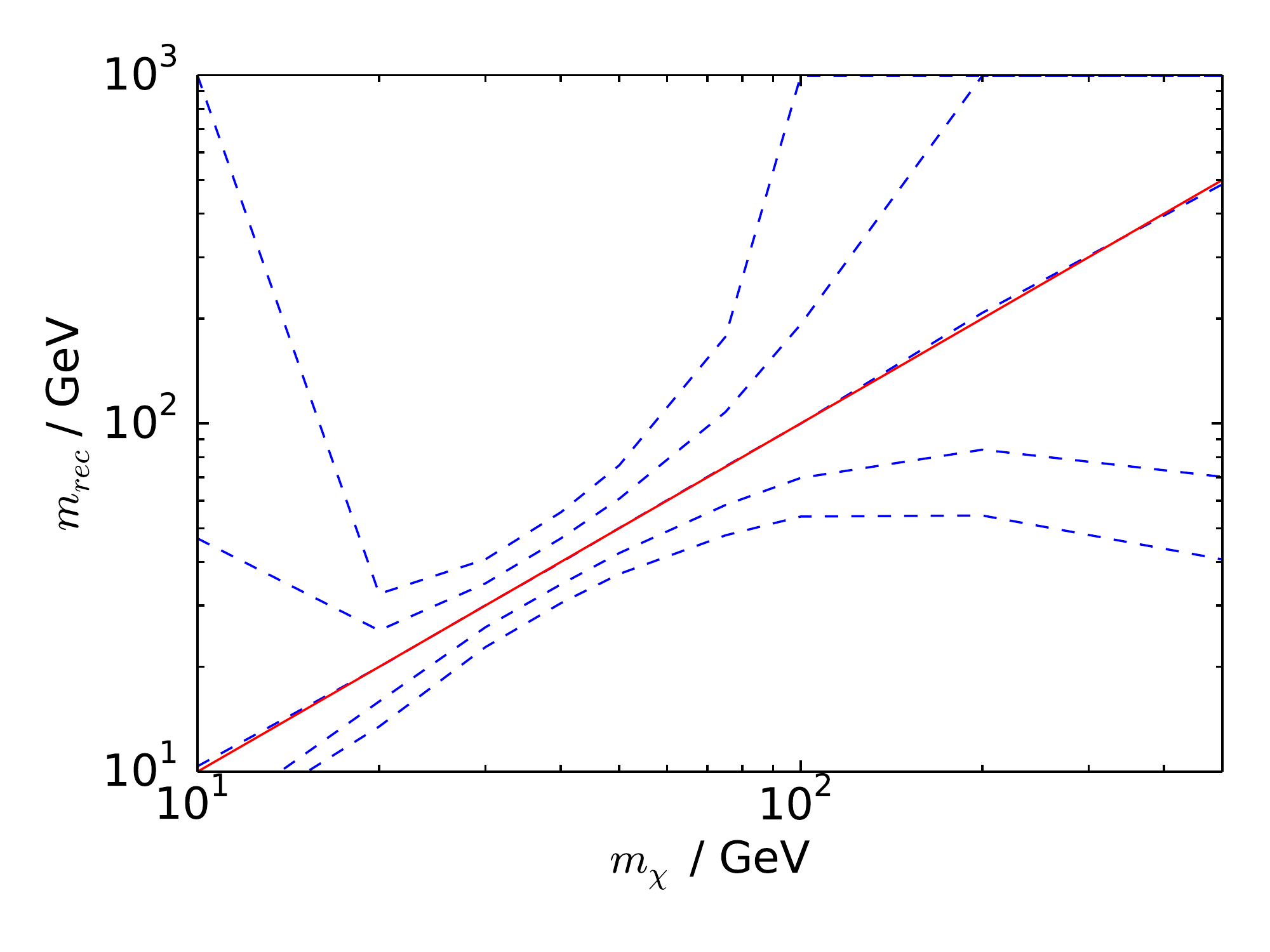}
  \caption{Reconstructed WIMP mass $m_\textrm{rec}$ (central dashed blue line) as a function of input WIMP mass $m_\chi$ as well as 68\% and 95\% intervals (inner and outer blue dashed lines respectively). The line $m_\textrm{rec} = m_\chi$ (solid red line) is also plotted for reference.}
  \label{fig:VaryingM}
\end{figure}

In previous work \cite{Kavanagh:2013a}, this parametrization method was only tested for a single WIMP mass of $50 \textrm{ GeV}$. Here, we extend this analysis to a wider range of WIMP masses. We generate Asimov data for WIMP masses of 10, 20, 30, 40, 50, 75, 100, 200 and 500 GeV and reconstruct the best fit WIMP mass $m_\textrm{rec}$ and 68\% and 95\% confidence intervals from the profile likelihood. We use the SHM as a benchmark distribution function and use a fixed number of $N=5$ basis functions. The results are shown in Fig.~\ref{fig:VaryingM}, along with the line $m_\textrm{rec} = m_\chi$ for reference.

For large values of $m_\chi$, the shape of the event spectrum becomes independent of $m_\chi$ \cite{Green:2008}, which results in a widening of the confidence intervals as the WIMP mass increases. For low mass WIMPs, fewer events are observed in each bin, again resulting in wider confidence intervals. It should be noted that for this analysis we have used Asimov data, in which the exact (non-integer) number of events is recorded in each bin. For low mass WIMPs, this means that the spectrum (and therefore the correct WIMP mass) is still well reconstructed using Asimov data, in spite of the small number of events. The tightest constraints are obtained when the input WIMP mass is close to the masses of several of the detector nuclei (in the range 30-80 GeV). There also appears to be no bias in the WIMP mass: the reconstruction matches the true mass across all values considered.

An alternative parametrization method was proposed in Ref.~\cite{Kavanagh:2012}, in which the \textit{momentum} distribution of halo WIMPs was parametrized. For a given speed distribution, the corresponding momentum distribution may be broad and easily reconstructed for high mass WIMPs. However, for low mass WIMPs the momentum distribution would be much narrower, owing to their lower momenta. The momentum parametrization method therefore performs poorly for low mass WIMPs. The parametrization presented in this paper does not suffer from similar problems.

So far, we have only considered idealized direct detection experiments. We now apply the method to more realistic mock detectors, taking into account the effects of finite energy resolution, as well as unrejected background events. We assume here that each experiment has a gaussian energy resolution with fixed width $\sigma_E = 1 \textrm{ keV}$, such that the observed event rate for recoils of energy $E$ is given by:

\begin{equation}
\frac{\textrm{d}R}{\textrm{d}E} = \int_{0}^{\infty} \frac{1}{\sqrt{2 \pi} \sigma_E}\exp\left\{-\frac{(E-E')^2}{2\sigma_E^2}\right\} \frac{\textrm{d}{R'}}{\textrm{d}E'} \, \textrm{d}E'\,,
\end{equation}
where the primed event rate is the underlying (perfect resolution) rate. We also assume a constant flat background rate for each experiment $R_\textrm{BG} = 10^{-6}$ events/kg/keV/day (which has been suggested as a possible background rate for Xenon1T \cite{Aprile:2010} and WArP-100L \cite{Grandi:2005}) when generating mock data sets. However, we allow the flat background rate in each experiment to vary as free parameters during the fit.

We have chosen relatively generic resolution and background parameters in this work, because the precise details of energy resolution and background shape and rate will depend on the specific experiment under consideration. Instead, we hope to show that the inclusion of more realistic experimental setups does not introduce an additional bias or otherwise spoil the good properties of the method presented here.

\begin{figure}[t]
  \includegraphics[width=0.49\textwidth]{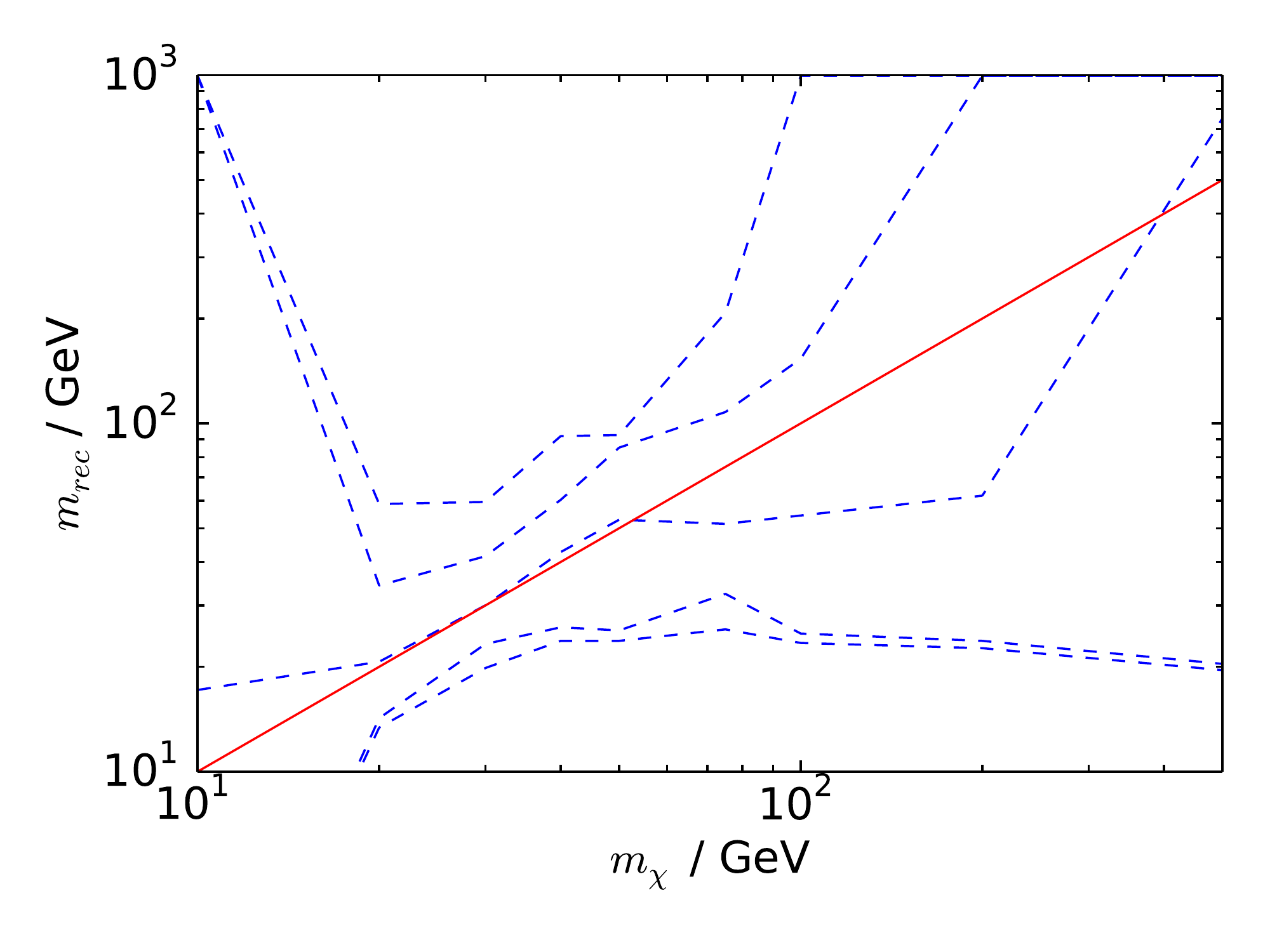}
  \caption{As fig.~\ref{fig:VaryingM} but including the effects of finite energy resolution and non-zero backgrounds, as described in the text.}
  \label{fig:VaryingM_real}
\end{figure}

Figure \ref{fig:VaryingM_real} shows the reconstructed mass as a function of input mass in this more realistic scenario. The 68\% and 95\% confidence intervals are now wider and the reconstructed mass does not appear to be as accurate. For input masses above $\sim$100 GeV, the uncertainties become very wide, with only a lower limit of $m_\textrm{rec} > 20 \textrm{ GeV}$ being placed on the WIMP mass.  Due to the poorer energy resolution the shape of the energy spectrum is less well-determined. In addition, a flat background contribution can mimic a higher mass WIMP, as it leads to a flatter spectrum. This leads to a strong degeneracy, as a wide range of mass values can provide a good fit to the data. For high input masses, the profile likelihood is approximately constant above $m_\textrm{rec} \sim 20 \textrm{ GeV}$, indicating that there is no sensitivity to the underlying WIMP mass.

In spite of this, the true mass values still lie within the 68\% and 95\% confidence intervals. In addition, the poor values for the reconstructed mass for heavy WIMPs are a side effect of the loss of sensitivity. Because the profile likelihood is approximately flat, the maximum likelihood point is equally likely to be anywhere within the 68\% interval. These effects would be present even if we had considered a fixed form for the speed distribution. However, when we allow for a range of possible speed distributions, the effects become more pronounced. These results show that for more realistic experimental scenarios, the method presented in this paper remains reliable over a range of masses, though its precision may be significantly reduced.

\section{Statistical properties}
\label{sec:stats}
We now consider the impact of statistical fluctuations on the reconstruction of the WIMP mass. In reality, the number of events observed $N_o$ at a given experiment will be Poisson distributed about the expected value $N_e$, while the observed distribution of recoil energies will not exactly match that expected from the calculated event rate. The fundamental statistical limitations of future direct detection experiments have been studied in detail in Ref.~\cite{Strege:2012}. In this work, we generate 250 realisations of data from the mock experiments described in Tab.~\ref{tab:experiments}. Each realisation of the mock data is generated as follows:

\begin{enumerate}
\item Calculate the number of expected events $N_e$, given $\left\{m_\chi, \sigma_p, f(v)\right\}$, using Eq.~\ref{eq:N_expected},
\item Pick the number of observed events $N_o$ from a Poisson distribution with mean $N_e$,
\item Pick recoil energies $\left\{E_1, E_2, ..., E_{N_o}\right\}$, from the distribution $P(E)$ in Eq.~\ref{eq:eventdistribution},
\item Repeat for all three experiments.
\end{enumerate}

\begin{figure}
  \includegraphics[width=0.49\textwidth]{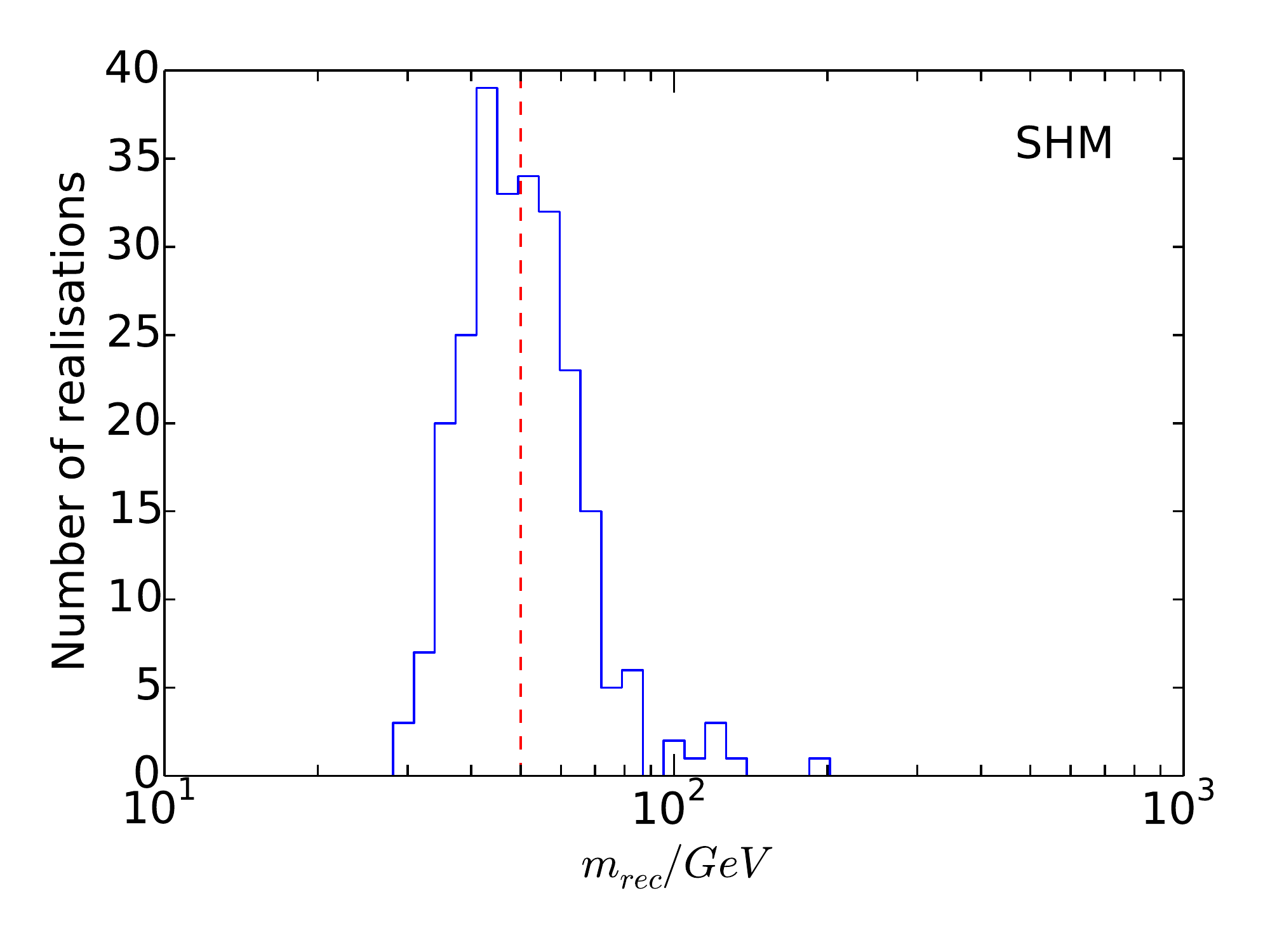}
  \includegraphics[width=0.49\textwidth]{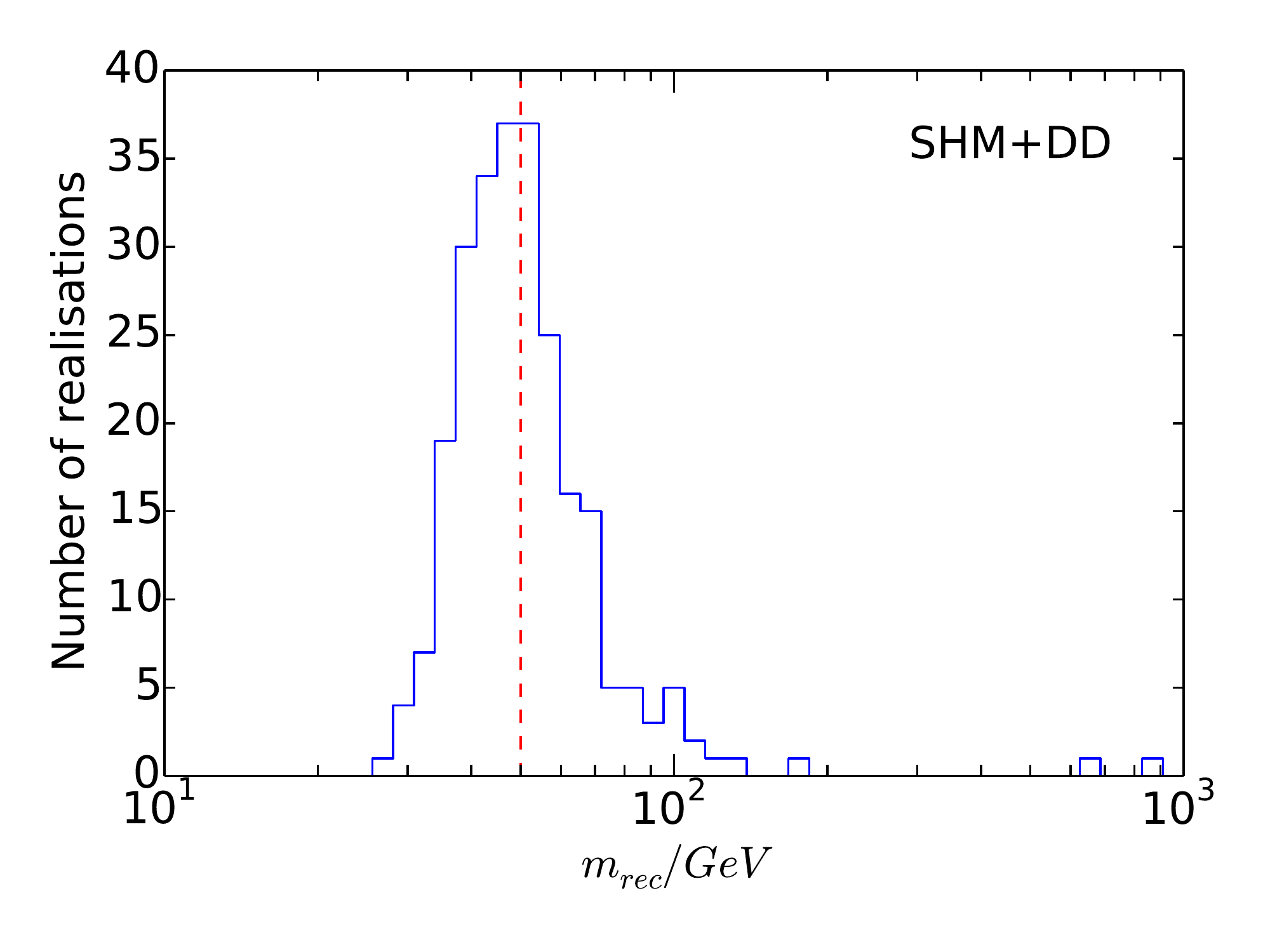}
  \includegraphics[width=0.49\textwidth]{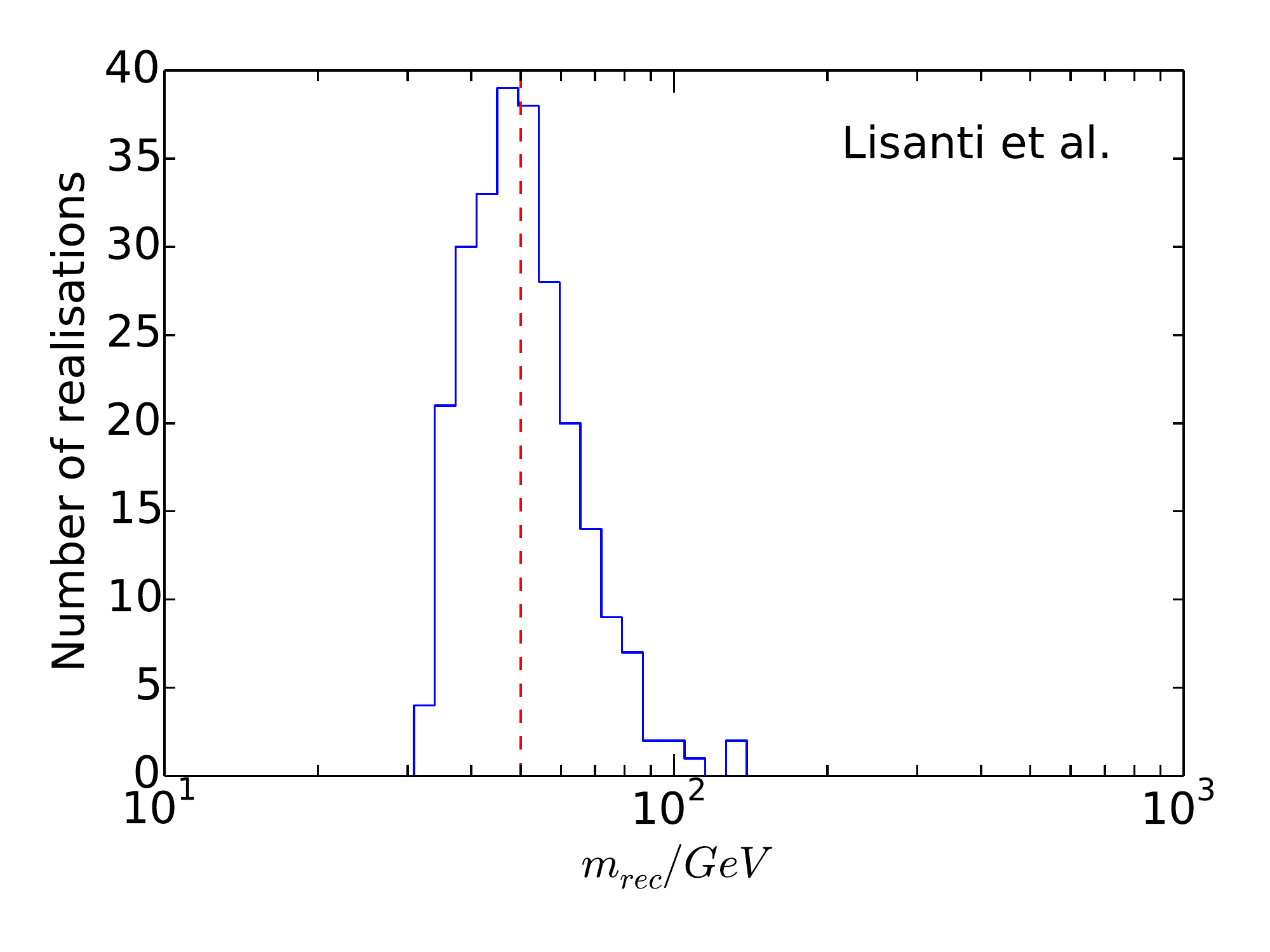}
  \caption{Distribution of the reconstructed mass $m_\textrm{rec}$ for 250 mock data sets generated using several benchmark speed distributions, defined in Sec.~\ref{sec:DDRate}. These are the SHM (top), SHM+DD (middle) and Lisanti et al.\ (bottom) distributions. The input WIMP mass of $m_\chi = 50 \textrm{ GeV}$ is shown as a vertical dashed red line.}
  \label{fig:Realisations}
\end{figure}

For each realisation, we then use the method described in Sec.~\ref{sec:ParameterRecon} (using $N = 5$ basis functions) to reconstruct the WIMP mass and 68\% and 95\% credible intervals. Figure~\ref{fig:Realisations} shows the distribution of reconstructed masses for an input mass of 50 GeV for three benchmark speed distributions: SHM, SHM+DD and Lisanti et al.\, as described in Sec.~\ref{sec:DDRate}. In all three cases, the reconstructions are peaked close to the true value, regardless of the underlying distribution. For the SHM+DD distribution, the spread of reconstructions is slightly wider (with more reconstructions extending up to higher masses). This is due to the smaller number of events for this benchmark, making the data sets more susceptible to Poisson fluctuations.

In order to assess the accuracy of the reconstructed value of the mass $m_\textrm{rec}$, we also calculate the bias $b$ for each realisation:

\begin{equation}
\label{eq:bias}
b = \textrm{ln}(m_\textrm{rec} / \textrm{GeV}) - \textrm{ln}(m_\textrm{true} / \textrm{GeV})\,.
\end{equation}
We compare the logarithms of the mass values because we have used logarithmically-flat priors on the WIMP mass. In Tab.~\ref{tab:bias} we show the average bias across all 250 realisations for each of the three benchmark distributions. In all three cases, the average bias is consistent with zero. Even in the SHM+DD case, which shows larger fluctuations away from the true value, there is no statistical bias.

\begin{table}[t]
  \setlength{\extrarowheight}{3pt}
  \setlength{\tabcolsep}{3pt}
  \begin{center}
	\begin{tabular}{m{1in}|c}
	Benchmark speed distribution & Mean bias $\langle b \rangle$ \\
	\hline\hline
	SHM & 0.002 $\pm$ 0.008 \\
	SHM+DD & 0.005 $\pm$ 0.007 \\
	Lisanti et al. & 0.01 $\pm$ 0.01 \\
	\end{tabular}
  \end{center}
\caption{\label{tab:bias} Mean bias $\langle b \rangle$ in the reconstructed log WIMP mass (Eq.~\ref{eq:bias}). This was calculated over 250 realisations using three different benchmark speed distributions.}
\end{table}

We also test the \textit{coverage} of the credible intervals which have been constructed. For a $p\%$ credible interval, we expect that the true parameter value of the WIMP mass will lie within the interval in $p\%$ of realisations. In this case, we say that the method provides \textit{exact coverage}. However, if the true parameter lies within the interval in fewer than $p\%$ of realisations, our reconstructed credible intervals are too narrow and provide \textit{undercoverage}. Alternatively, we obtain \textit{overcoverage} when the true parameter lies within the interval more often that $p\%$ of the time. Table \ref{tab:coverage} shows the coverage values for the $68\%$ and $95\%$ intervals obtained in this section. In each case, there is very close to exact coverage. We have also checked that these intervals only provide exact coverage for the true WIMP mass of 50 GeV. Other values of $m_\textrm{rec}$ are contained within the intervals less frequently than the true value, again indicating that this parametrization allows for unbiased and statistically robust reconstructions of the WIMP mass.

\begin{table}[t]
  \setlength{\extrarowheight}{3pt}
  \setlength{\tabcolsep}{3pt}
  \begin{center}
	\begin{tabular}{m{1in}|cc}
	Benchmark speed distribution & 68\% coverage & 95\% coverage\\
	\hline\hline
	SHM &  71 $\pm$ 3 \% & 94 $\pm$ 3 \%  \\
	SHM+DD & 68 $\pm$ 3 \% & 91 $\pm$ 4 \%  \\
	Lisanti et al. & 70 $\pm$ 3 \% & 95 $\pm$ 3 \%  \\
	\end{tabular}
  \end{center}
\caption{Coverage of 68\% and 95\% credible intervals calculated from 250 data realisations each for three benchmark speed distributions. The concept of coverage is described in the text of Sec.~\ref{sec:stats}.}
\label{tab:coverage}
\end{table}

\section{Reconstructing $f_1(v)$}
\label{sec:Recon}

Using the method described in this paper, we can obtain the posterior probability distribution for the coefficients $\left\{ a_1, ..., a_{N-1}\right\}$ given the data, which we refer to as $P(\textbf{a})$. We would like to be able to present this information in terms of the distribution function $f_1(v)$ in order to compare with some known distribution or look for particular features in the distribution. However, due to the fact that the distribution function is normalized, the values of $f_1$ at different speeds will be strongly correlated. We illustrate here how robust comparisons with benchmark distributions can be made.

\begin{figure}[t]
  \includegraphics[width=0.49\textwidth]{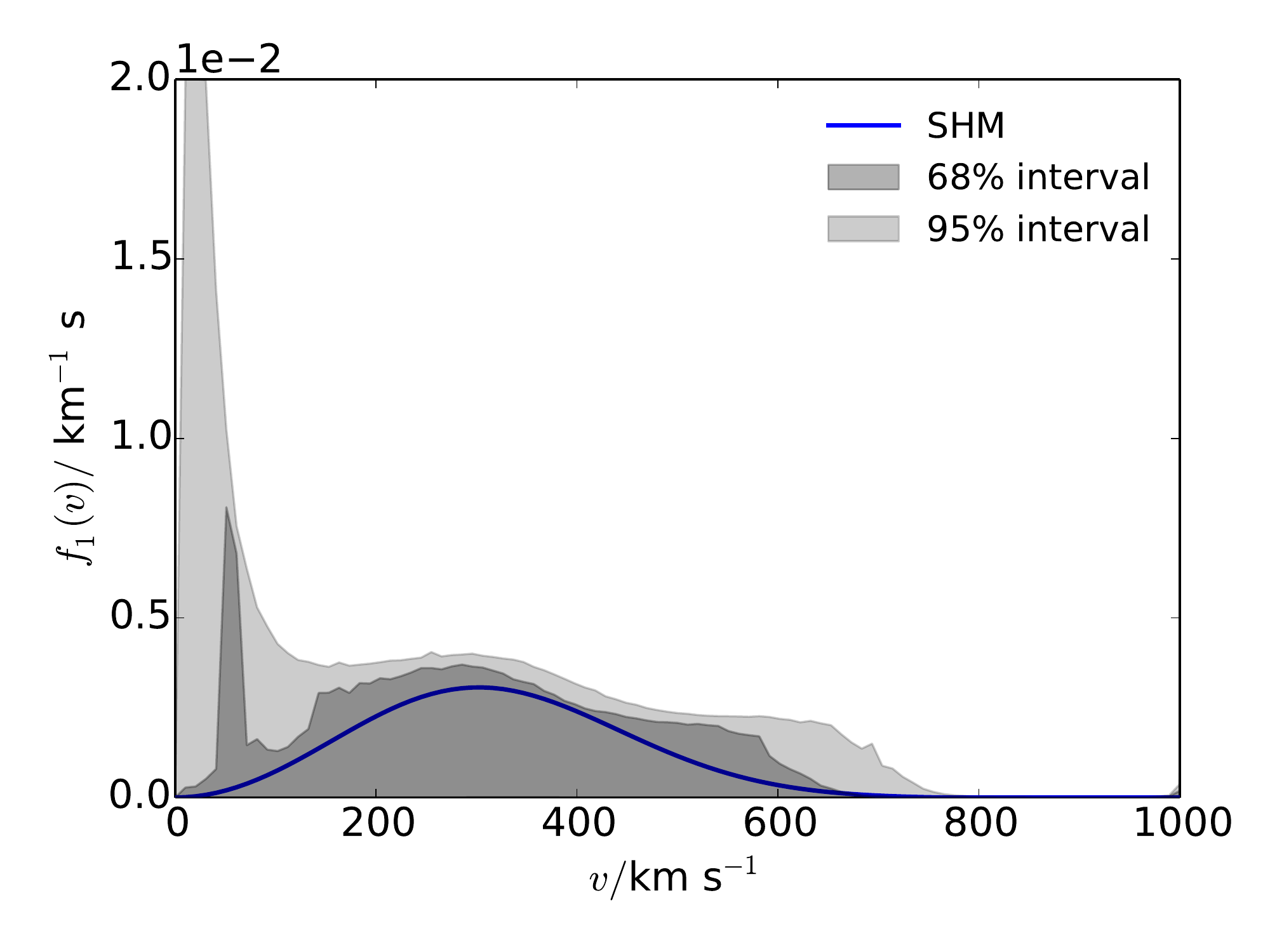}
  \caption{Reconstructed speed distribution for a single realisation of data, generated for a 50 GeV WIMP. 68\% and 95\% credible intervals are shown as dark and light shaded regions respectively, while the underlying SHM distribution function is shown as a solid blue line.}
  \label{fig:f}
\end{figure}

As a first step, we can attempt to sample from the $P(\textbf{a})$, in order to obtain $P(f_1(v))$. This is the probability distribution for the value of $f_1$ at a particular speed $v$, marginalizing over the values of $f_1$ at all other speeds. We can repeat for a range of speeds to obtain 68\% and 95\% credible intervals for the whole of $f_1(v)$. The result of this procedure is presented in Fig.~\ref{fig:f}, for a randomly selected realisation from the SHM ensemble of Sec.~\ref{sec:stats}. The underlying SHM distribution is shown as a solid line, while the 68\% and 95\% marginalized intervals are shown as dark and light shaded regions respectively. In this naive approach, we see that there is little shape information which can be recovered from the reconstruction, with only upper limits being placed on the speed distribution.

\begin{figure}[t]
  \includegraphics[width=0.49\textwidth]{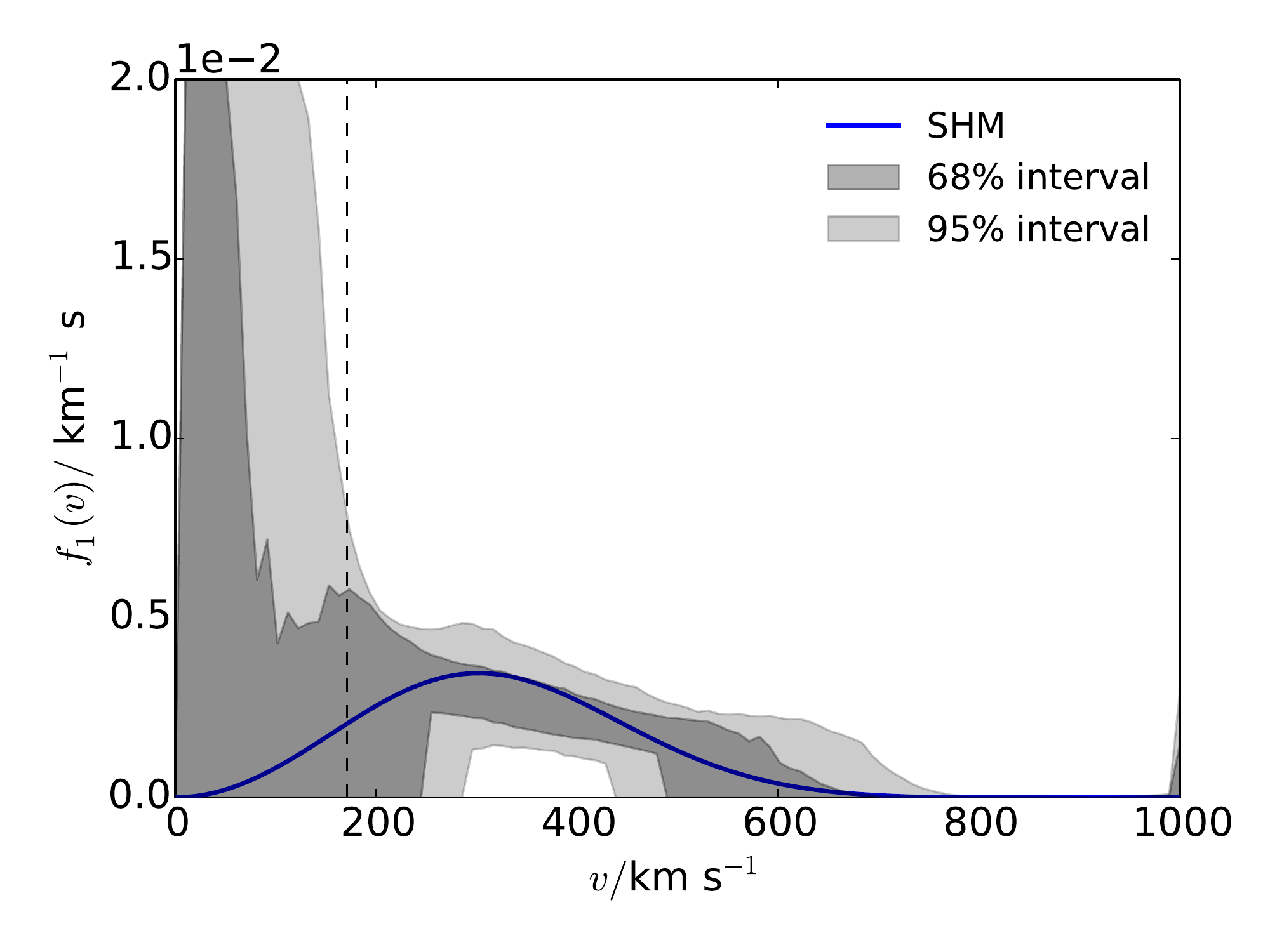}
  \caption{Reconstructed speed distribution for the same realisation of data as Fig.~\ref{fig:f}. In this case, we have also normalized $f_1(v)$ to unity above $v_a \approx 171 \kms$ (vertical dashed line). This is the lowest speed accessible to the experiments for a WIMP of mass 50 GeV. 68\% and 95\% credible intervals are shown as dark and light shaded regions respectively, while the underlying SHM distribution function is shown as a solid blue line.}
  \label{fig:f_scaled}
\end{figure}

This method performs poorly because, as initially mentioned in Sec.~\ref{sec:ParameterRecon}, we have no information about the fraction of dark matter particles below the energy threshold of our experiments. If this fraction is large, the event rate for a given cross-section is suppressed. However, increasing the cross-section will increase the total event rate. There is thus a degeneracy between the shape of the speed distribution and the cross-section, meaning that we can only probe the shape of $f_1(v)$, rather than its overall normalization. This degeneracy has not been accounted for in Fig.~\ref{fig:f}. We can attempt to correct for this by adjusting the normalization of $f_1(v)$. If we fix $f_1(v)$ to be normalized to unity above $v_a$ (where $v_a \approx 171 \kms$ is the lowest speed probed by the experiments for a WIMP of mass 50 GeV), we can compare the shapes of the underlying and reconstructed distribution functions. This is illustrated in Fig.~\ref{fig:f_scaled}, which shows that we now broadly reconstruct the correct shape of $f_1(v)$. Below $v_a$, the value of $f_1(v)$ is poorly constrained, because the experiments provide no information about the shape of the distribution below theshold.

There remain several issues with this approach. In order to utilize this method, we must know the approximate value of the lowest speed probed by the experiments. However, this value is set by the WIMP mass. We could determine $v_a$ using the reconstructed WIMP mass, but this would be subject to significant uncertainty. In addition, direct reconstructions of the speed distribution are easily biased. The upper limit of the energy windows of the experiments corresponds to a particular WIMP speed (for a given WIMP mass). WIMPs above this speed still contribute to the total event rate, but contribute no spectral information. The reconstructed shape of the high speed tail of the distribution is therefore not constrained by the data, but may affect the reconstructed value of $f_1$ at lower speeds.

An alternative approach is to reconstruct the mean inverse speed $\eta(v)$ (defined in Eq.~\ref{eq:eta}) at some speed $v$. Because $\eta(v)$ is an integral function of $f_1$, it is less prone to bias as it takes into account the full shape of the distribution at speeds greater than $v$. However, we do not know the normalization of $f_1$ and so we must normalize $\eta$ appropriately. For each point sampled from $P(\textbf{a})$, we calculate $\eta$. We then divide by $\alpha(v)$, the fraction of WIMPs above speed $v$, calculated using the same parameter point:
\begin{equation}
\label{eq:alpha}
\alpha(v) = \int_{v}^{\infty} f_1(v') \, \textrm{d}v'\,.
\end{equation}

We will write this rescaled mean inverse speed as $\eta^*(v) = \eta(v)/\alpha(v)$. The value of $\eta^*(v)$ is a measure of the shape of the distribution function above $v$. However, information about the normalization of the distribution has been factored out by dividing by $\alpha(v)$. We no longer need to know the value of $v_a$ in order to obtain information about the shape of the distribution at higher speeds. We may still need to decide the speed down to which we trust our reconstruction, but this no longer relies on an arbitrary choice of $v_a$ to normalize the reconstructions at all speeds.

\begin{figure}[t]
  \includegraphics[width=0.49\textwidth]{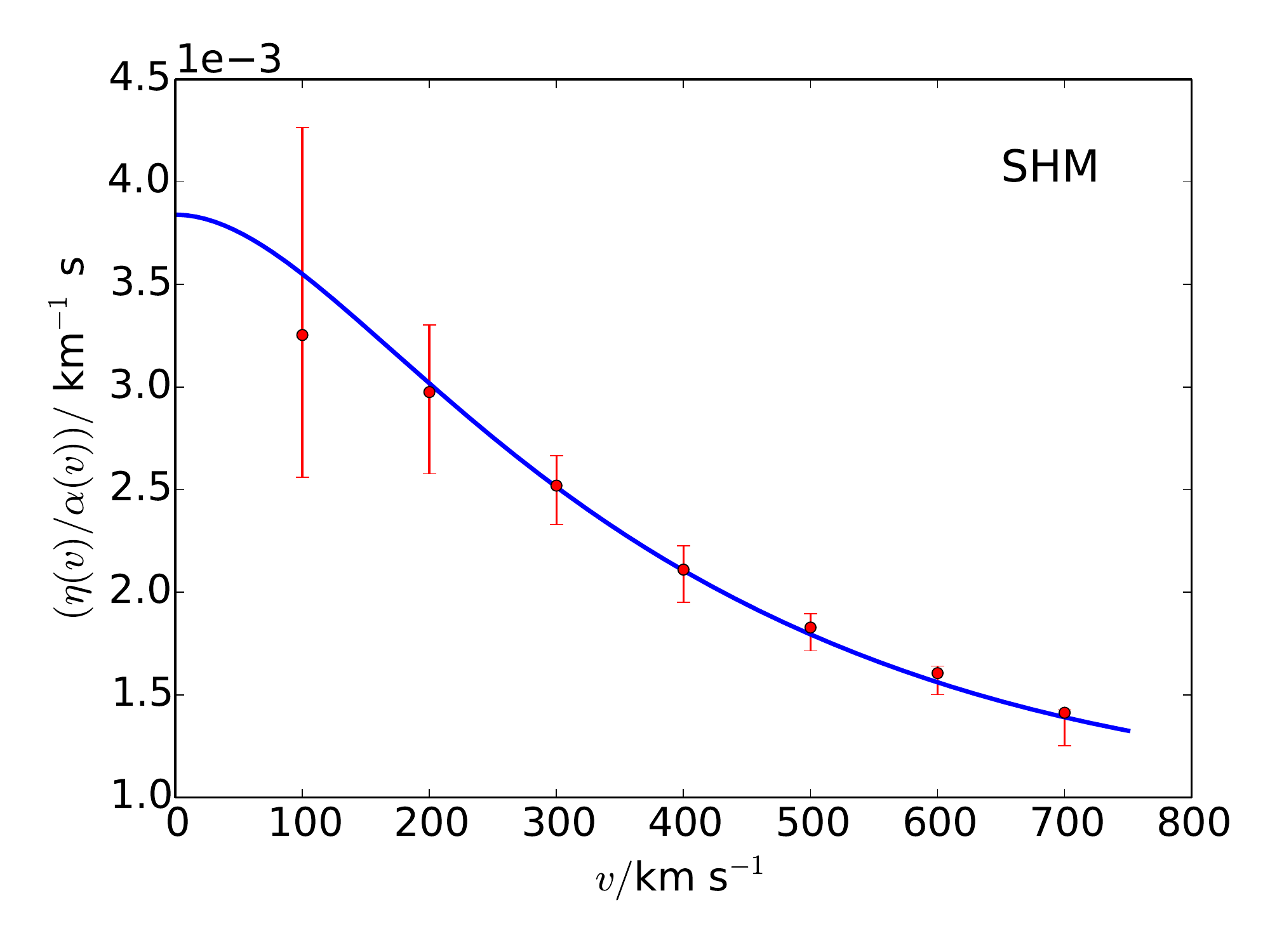}
  \caption{Mean reconstructed values of the rescaled mean inverse speed $\eta(v)/\alpha(v)$ at several values of $v$, calculated over 250 realisations of data using a 50 GeV WIMP and underlying SHM distribution function. Errorbars indicate the mean upper and lower limits of the 68\% credible intervals. The underlying form of $\eta(v)/\alpha(v)$ obtained from the SHM is shown as a solid blue line.}
  \label{fig:eta_stats}
\end{figure}

In Fig.~\ref{fig:eta_stats}, we plot the mean reconstructed value of $\eta^*$ at several values of $v$, using 250 realisations of the 50 GeV SHM benchmark. We also show the mean upper and lower limits of the 68\% credible intervals as errorbars. The form of $\eta^*$ for the SHM is shown as a solid blue line. In all cases except for $v=100 \kms$, the mean reconstructed value is close to the true value, indicating that $\eta^*$ can be reconstructed without bias using this method. At low speeds, the reconstructed value deviates from the true value. In addition, the credible intervals lead to \textit{under}coverage in the $v=100 \kms$ case. However, this point lies below the lowest speed to which the experiments are sensitive and therefore we cannot trust the reconstruction at this low speed. We have checked that for the remaining values of $v$ the method provides exact or overcoverage, indicating that at higher speeds we can use $\eta^*$ as a reliable and statistically robust measure of the shape of the distribution.

In the case of a single realisation of data, we would like to compare the probability distribution for $\eta^*(v)$ (obtained from $P(\textbf{a})$) to the value calculated from some test distribution. We note that several distributions may produce the same value of $\eta^*(v)$ at a given value of $v$. Thus, we may fail to reject a distribution function which is not the true distribution. However, if the calculated value of $\eta^*(v)$ does lie outside the $p\%$ interval, we can reject it at the $p\%$ level.

\begin{figure}[t]
  \includegraphics[width=0.49\textwidth]{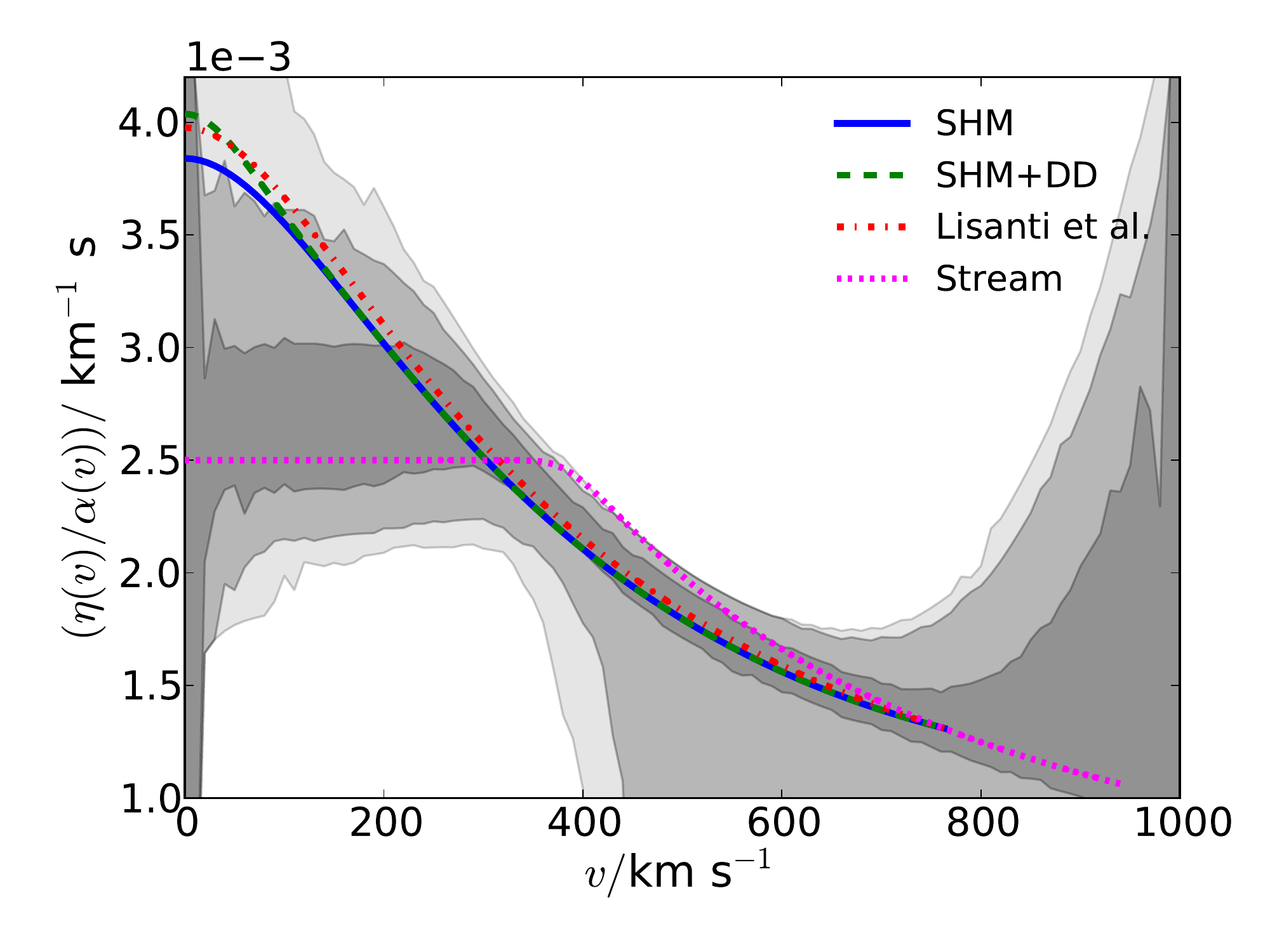}
  \caption{Rescaled mean inverse speed $\eta(v)/\alpha(v)$, reconstructed from a single realisation of data using a 50 GeV WIMP and underlying SHM distribution function. At each value of $v$ we calculate 68\%, 95\% and 99\% credible intervals (shown as shaded intervals). We also show the calculated values of $\eta(v)/\alpha(v)$ for several possible benchmark speed distributions: SHM (solid blue), SHM+DD (dashed green), Lisanti et al.\ (dot-dashed red) and stream (dotted magenta). The benchmark curves are truncated when the underlying distribution function goes to zero.}
  \label{fig:eta}
\end{figure}

We can increase the discriminating power of this method by repeating this reconstruction over all speeds and checking to see if the benchmark value of $\eta^*$ is rejected at any value of $v$. The result of this procedure is shown in Fig.~\ref{fig:eta} for a single realisation of data generated using an SHM distribution (the same data as in Figs. \ref{fig:f} and \ref{fig:f_scaled}). We plot the 68\%, 95\% and 99\% credible intervals as shaded regions, as well as the values of $\eta^*(v)$ calculated from several benchmark speed distribution. We will focus on the intermediate speed range ($v \gtrsim 200 \kms$), as we do not know \textit{a priori} the lowest speed to which the experiments are sensitive.

The reconstructed intervals are consistent with a range of possible distribution functions. The SHM and SHM+DD distributions are identical over a wide range of speeds. This is because above $\sim 200 \kms$, the two distributions differ in normalization but not in shape. Differences appear between the two at low speeds where their shapes diverge. The Lisanti et al.\ distribution results in a larger deviation from the SHM, but not sufficiently large to differentiate between the two distributions given the size of the uncertainties. Finally, the stream distribution results in a significantly different form for $\eta^*(v)$. At approximately $400 \kms$, the curve for the stream distribution lies outside the reconstructed 99\% credible interval. We can therefore use this method to reject the stream distribution at the 99\% confidence level.

\begin{figure}[t]
  \includegraphics[width=0.49\textwidth]{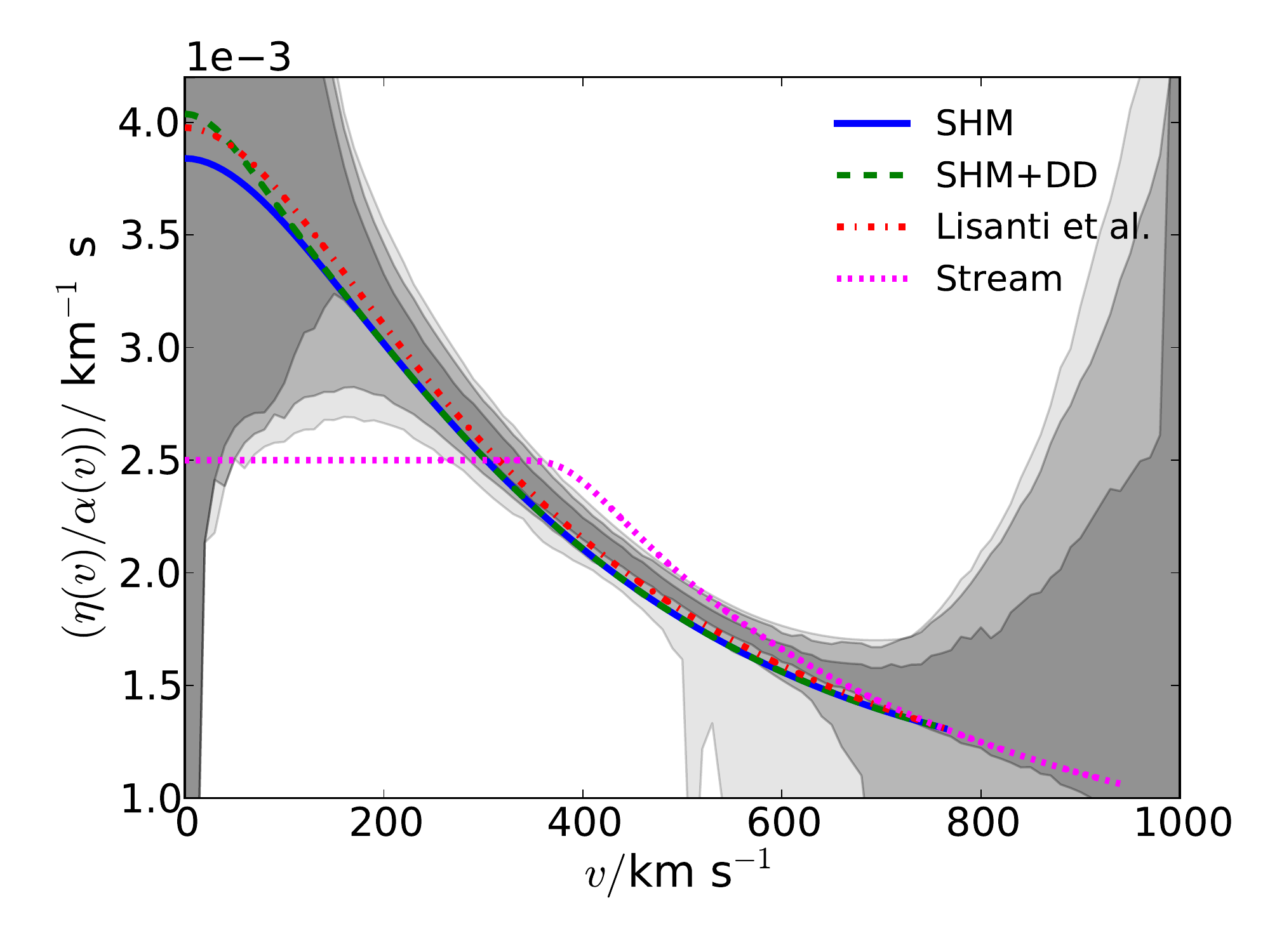}
  \caption{As Fig.~\ref{fig:eta}, but using as input a Lisanti et al.\ speed distribution and an exposure time which is 2.5 times longer.}
  \label{fig:eta_hires}
\end{figure}

\begin{figure}[t]
  \includegraphics[width=0.49\textwidth]{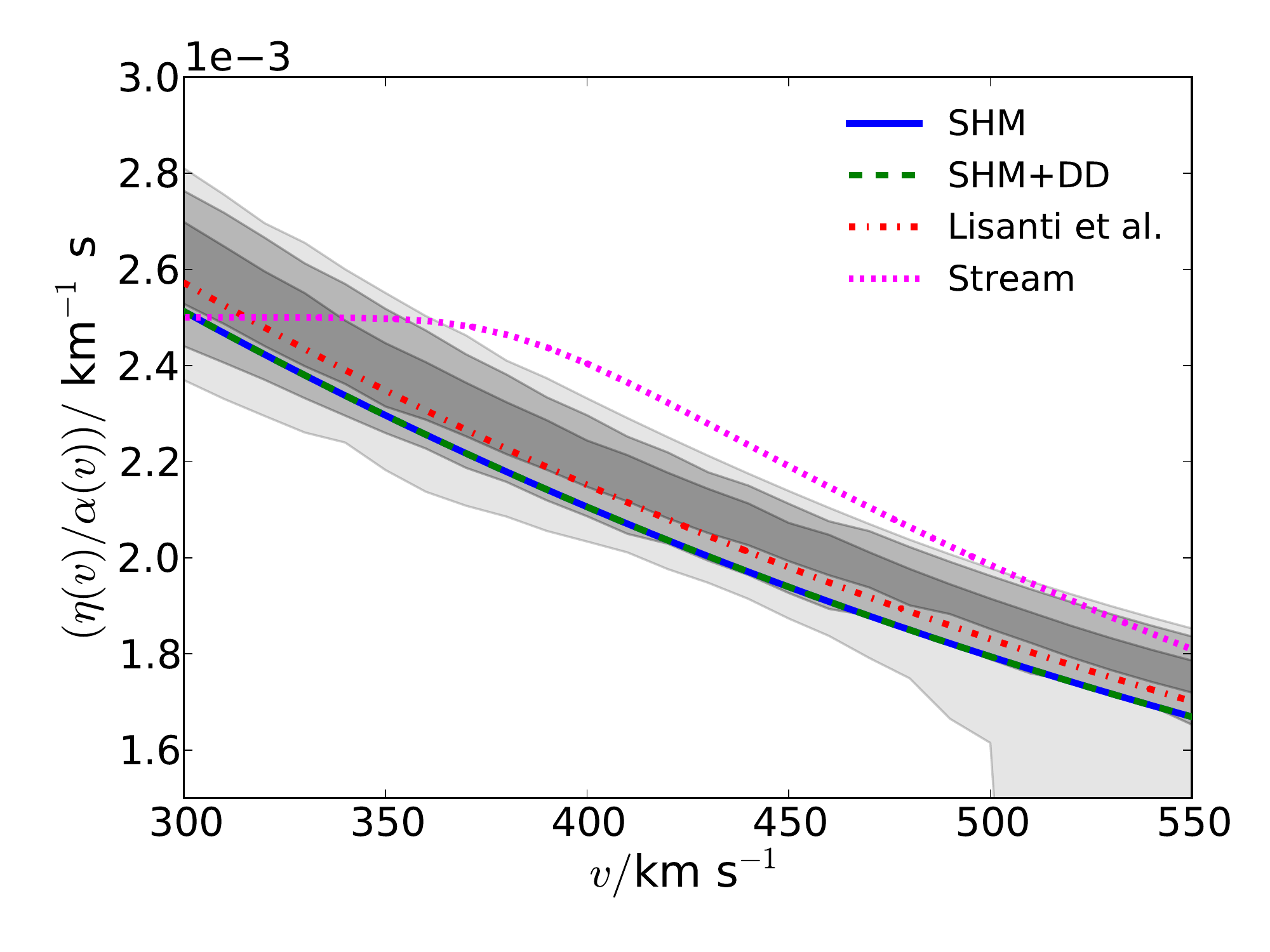}
  \caption{As Fig.~\ref{fig:eta_hires}, but focusing on the region around $v \sim 400 \kms$. Notice that in the range $400-550 \kms$, both the SHM and SHM+DD curves lie at or below the lower limit of the 95\% credible interval.}
  \label{fig:eta_hires_zoom}
\end{figure}

Figure \ref{fig:eta_hires} shows the results of a reconstruction using a larger exposure. In this case, we generate data using the Lisanti et al.\ distribution and an exposure increased by a factor of $2.5$, resulting in approximately 1000 events across the three detectors. As expected, the resulting credible intervals are now substantially narrower. The stream distribution now lies significantly outside the 99\% interval. In Fig.~\ref{fig:eta_hires_zoom}, we show the same results, but focusing in on the region around $v \sim 400 \kms$. At certain points, the SHM and SHM+DD distributions now lie outside the 95\% credible interval, suggesting that with a number of events of the order of 1000, we may be able to reject these benchmarks.

While the method displayed in Fig.~\ref{fig:f_scaled} allows the approximate shape of the speed distribution to be reconstructed, reconstructions of $\eta^*(v)$ allow more statistically robust statements to be made about the underlying speed distribution. In particular, Fig.~\ref{fig:eta_hires_zoom} illustrates that with larger exposures deviations from Maxwellian speed distributions can be detected in a model-independent fashion.

\section{Conclusions}
\label{sec:Conclusions}

We have studied in detail the parametrization for the local dark matter speed distribution introduced in Paper 1. This method involves writing the logarithm of the speed distribution as a polynomial in speed $v$ and fitting the polynomial coefficients (along with the WIMP mass and cross section) to the data. We have attempted to disentangle in this paper the influence of  different benchmark speed distributions, different benchmark WIMP masses and different forms for the parametrization. We summarize our conclusions as follows:

\begin{itemize}

\item We have shown that the reconstruction of the WIMP mass is robust under changes in the number of basis functions $N$. We have used the Bayesian Information Criterion (BIC) to compare models with different values of $N$ and have shown that minimizing the BIC allows us to determine how many basis functions are required for a reliable reconstruction. We have also demonstrated that the results of the method do not depend strongly on the choice of basis functions, but that the speed of reconstructions may improved by using the Chebyshev polynomial basis.

\item We have shown that the method leads to unbiased reconstructions of the WIMP mass for masses in the range 10-500 GeV. Including realistic experimental parameters, including non-zero backgrounds and finite energy resolution, reduces the precision of these reconstructions. In particular, for large values of the input mass, we can only place a lower limit of approximately 20 GeV on the reconstructed mass. This is significantly lower than in the idealized case, where we can typically constrain the WIMP mass to be heavier than around 50 GeV.

\item We have used several ensembles of data realisations to demonstrate the statistical properties of the method, including unbiased reconstructions and exact coverage of the WIMP mass.

\item We have presented several ways of displaying the reconstructed WIMP speed distribution using this method. In order to make robust statistical inferences about the speed distribution, we calculate the probability distribution of $\eta(v)/\alpha(v)$. This is the mean inverse speed $\eta(v)$, which appears in the direct detection event rate (eq.~\ref{eq:Rate}), rescaled by the fraction of WIMPs $\alpha(v)$ above speed $v$. This can be used as a measure of the \textit{shape} of the distribution function, from which the unknown normalization has been factored out. We can then compare to the expected value of $\eta(v)/\alpha(v)$ from a given benchmark speed distribution, allowing us to distinguish between different underlying models.

\end{itemize}

We have shown that this parametrization method is statistically robust and works well over a large range of input parameters, both in terms of particle physics and astrophysics. The inclusion of more realistic experimental parameters does not introduce any additional bias, but does reduce the precision of reconstructions. We obtain unbiased estimates of the WIMP mass over large numbers of data sets. Finally, we have shown that we can distinguish different forms of the speed distribution. With around 1000 events, it may be possible to detect minor deviations from the Standard Halo Model and begin to search for more interesting structure in the speed distribution of the Milky Way.

\begin{acknowledgments}
 The author thanks Anne M. Green and Mattia Fornasa for helpful comments. BJK is supported by STFC. Access to the University of Nottingham High Performance Computing Facility is also gratefully acknowledged.
\end{acknowledgments}

\bibliographystyle{apsrev4-1}
\bibliography{Model_Indep}

\begin{thebibliography}{64}%
\makeatletter
\providecommand \@ifxundefined [1]{%
 \@ifx{#1\undefined}
}%
\providecommand \@ifnum [1]{%
 \ifnum #1\expandafter \@firstoftwo
 \else \expandafter \@secondoftwo
 \fi
}%
\providecommand \@ifx [1]{%
 \ifx #1\expandafter \@firstoftwo
 \else \expandafter \@secondoftwo
 \fi
}%
\providecommand \natexlab [1]{#1}%
\providecommand \enquote  [1]{``#1''}%
\providecommand \bibnamefont  [1]{#1}%
\providecommand \bibfnamefont [1]{#1}%
\providecommand \citenamefont [1]{#1}%
\providecommand \href@noop [0]{\@secondoftwo}%
\providecommand \href [0]{\begingroup \@sanitize@url \@href}%
\providecommand \@href[1]{\@@startlink{#1}\@@href}%
\providecommand \@@href[1]{\endgroup#1\@@endlink}%
\providecommand \@sanitize@url [0]{\catcode `\\12\catcode `\$12\catcode
  `\&12\catcode `\#12\catcode `\^12\catcode `\_12\catcode `\%12\relax}%
\providecommand \@@startlink[1]{}%
\providecommand \@@endlink[0]{}%
\providecommand \url  [0]{\begingroup\@sanitize@url \@url }%
\providecommand \@url [1]{\endgroup\@href {#1}{\urlprefix }}%
\providecommand \urlprefix  [0]{URL }%
\providecommand \Eprint [0]{\href }%
\providecommand \doibase [0]{http://dx.doi.org/}%
\providecommand \selectlanguage [0]{\@gobble}%
\providecommand \bibinfo  [0]{\@secondoftwo}%
\providecommand \bibfield  [0]{\@secondoftwo}%
\providecommand \translation [1]{[#1]}%
\providecommand \BibitemOpen [0]{}%
\providecommand \bibitemStop [0]{}%
\providecommand \bibitemNoStop [0]{.\EOS\space}%
\providecommand \EOS [0]{\spacefactor3000\relax}%
\providecommand \BibitemShut  [1]{\csname bibitem#1\endcsname}%
\let\auto@bib@innerbib\@empty
\bibitem [{\citenamefont {Bertone}\ \emph {et~al.}(2005)\citenamefont
  {Bertone}, \citenamefont {Hooper},\ and\ \citenamefont
  {Silk}}]{Bertone:2005}%
  \BibitemOpen
  \bibfield  {author} {\bibinfo {author} {\bibfnamefont {G.}~\bibnamefont
  {Bertone}}, \bibinfo {author} {\bibfnamefont {D.}~\bibnamefont {Hooper}}, \
  and\ \bibinfo {author} {\bibfnamefont {J.}~\bibnamefont {Silk}},\ }\href
  {\doibase 10.1016/j.physrep.2004.08.031} {\bibfield  {journal} {\bibinfo
  {journal} {Phys. Rep.}\ }\textbf {\bibinfo {volume} {405}},\ \bibinfo {pages}
  {279} (\bibinfo {year} {2005})},\ \Eprint
  {http://arxiv.org/abs/hep-ph/0404175} {hep-ph/0404175} \BibitemShut {NoStop}%
\bibitem [{\citenamefont {Jungman}\ \emph {et~al.}(1996)\citenamefont
  {Jungman}, \citenamefont {Kamionkowski},\ and\ \citenamefont
  {Griest}}]{Jungman:1996}%
  \BibitemOpen
  \bibfield  {author} {\bibinfo {author} {\bibfnamefont {G.}~\bibnamefont
  {Jungman}}, \bibinfo {author} {\bibfnamefont {M.}~\bibnamefont
  {Kamionkowski}}, \ and\ \bibinfo {author} {\bibfnamefont {K.}~\bibnamefont
  {Griest}},\ }\href {\doibase 10.1016/0370-1573(95)00058-5} {\bibfield
  {journal} {\bibinfo  {journal} {Phys. Rep.}\ }\textbf {\bibinfo {volume}
  {267}},\ \bibinfo {pages} {195} (\bibinfo {year} {1996})},\ \Eprint
  {http://arxiv.org/abs/hep-ph/9506380} {hep-ph/9506380} \BibitemShut {NoStop}%
\bibitem [{\citenamefont {Dodelson}\ and\ \citenamefont
  {Widrow}(1994)}]{Dodelson:1994}%
  \BibitemOpen
  \bibfield  {author} {\bibinfo {author} {\bibfnamefont {S.}~\bibnamefont
  {Dodelson}}\ and\ \bibinfo {author} {\bibfnamefont {L.}~\bibnamefont
  {Widrow}},\ }\href {\doibase 10.1103/PhysRevLett.72.17} {\bibfield  {journal}
  {\bibinfo  {journal} {Phys. Rev. Lett.}\ }\textbf {\bibinfo {volume} {72}},\
  \bibinfo {pages} {17} (\bibinfo {year} {1994})},\ \Eprint
  {http://arxiv.org/abs/hep-ph/9303287} {hep-ph/9303287} \BibitemShut {NoStop}%
\bibitem [{\citenamefont {Duffy}\ and\ \citenamefont
  {Bibber}(2009)}]{Duffy:2009}%
  \BibitemOpen
  \bibfield  {author} {\bibinfo {author} {\bibfnamefont {L.~D.}\ \bibnamefont
  {Duffy}}\ and\ \bibinfo {author} {\bibfnamefont {K.~v.}\ \bibnamefont
  {Bibber}},\ }\href {\doibase 10.1088/1367-2630/11/10/105008} {\bibfield
  {journal} {\bibinfo  {journal} {New J. Phys.}\ }\textbf {\bibinfo {volume}
  {11}},\ \bibinfo {pages} {105008} (\bibinfo {year} {2009})},\ \Eprint
  {http://arxiv.org/abs/0904.3346} {arXiv:0904.3346} \BibitemShut {NoStop}%
\bibitem [{\citenamefont {Kolb}\ and\ \citenamefont
  {Slansky}(1984)}]{Kolb:1984}%
  \BibitemOpen
  \bibfield  {author} {\bibinfo {author} {\bibfnamefont {E.~W.}\ \bibnamefont
  {Kolb}}\ and\ \bibinfo {author} {\bibfnamefont {R.}~\bibnamefont {Slansky}},\
  }\href {\doibase 10.1016/0370-2693(84)90298-3} {\bibfield  {journal}
  {\bibinfo  {journal} {Phys. Lett. B}\ }\textbf {\bibinfo {volume} {135}},\
  \bibinfo {pages} {378} (\bibinfo {year} {1984})}\BibitemShut {NoStop}%
\bibitem [{\citenamefont {Goodman}\ and\ \citenamefont
  {Witten}(1985)}]{Goodman:1985}%
  \BibitemOpen
  \bibfield  {author} {\bibinfo {author} {\bibfnamefont {M.~W.}\ \bibnamefont
  {Goodman}}\ and\ \bibinfo {author} {\bibfnamefont {E.}~\bibnamefont
  {Witten}},\ }\href {\doibase 10.1103/physrevd.31.3059} {\bibfield  {journal}
  {\bibinfo  {journal} {Phys. Rev. D}\ }\textbf {\bibinfo {volume} {31}},\
  \bibinfo {pages} {3059} (\bibinfo {year} {1985})}\BibitemShut {NoStop}%
\bibitem [{\citenamefont {Drukier}\ \emph {et~al.}(1986)\citenamefont
  {Drukier}, \citenamefont {Freese},\ and\ \citenamefont
  {Spergel}}]{Drukier:1986}%
  \BibitemOpen
  \bibfield  {author} {\bibinfo {author} {\bibfnamefont {A.~K.}\ \bibnamefont
  {Drukier}}, \bibinfo {author} {\bibfnamefont {K.}~\bibnamefont {Freese}}, \
  and\ \bibinfo {author} {\bibfnamefont {D.~N.}\ \bibnamefont {Spergel}},\
  }\href {\doibase 10.1103/physrevd.33.3495} {\bibfield  {journal} {\bibinfo
  {journal} {Phys. Rev. D}\ }\textbf {\bibinfo {volume} {33}},\ \bibinfo
  {pages} {3495} (\bibinfo {year} {1986})}\BibitemShut {NoStop}%
\bibitem [{\citenamefont {Lavalle}\ and\ \citenamefont
  {Salati}(2012)}]{Lavalle:2012}%
  \BibitemOpen
  \bibfield  {author} {\bibinfo {author} {\bibfnamefont {J.}~\bibnamefont
  {Lavalle}}\ and\ \bibinfo {author} {\bibfnamefont {P.}~\bibnamefont
  {Salati}},\ }\href {\doibase 10.1016/j.crhy.2012.05.001} {\bibfield
  {journal} {\bibinfo  {journal} {C. R. Phys.}\ }\textbf {\bibinfo {volume}
  {13}},\ \bibinfo {pages} {740} (\bibinfo {year} {2012})},\ \Eprint
  {http://arxiv.org/abs/1205.1004} {arXiv:1205.1004} \BibitemShut {NoStop}%
\bibitem [{\citenamefont {Battaglia}\ and\ \citenamefont
  {Peskin}(2010)}]{Battaglia:2010}%
  \BibitemOpen
  \bibfield  {author} {\bibinfo {author} {\bibfnamefont {M.}~\bibnamefont
  {Battaglia}}\ and\ \bibinfo {author} {\bibfnamefont {M.~E.}\ \bibnamefont
  {Peskin}},\ }in\ \href@noop {} {\emph {\bibinfo {booktitle} {Particle Dark
  Matter: Observation, Models and Searches}}},\ \bibinfo {editor} {edited by\
  \bibinfo {editor} {\bibfnamefont {G.}~\bibnamefont {Bertone}}}\ (\bibinfo
  {publisher} {Cambridge University Press},\ \bibinfo {year} {2010})\
  Chap.~\bibinfo {chapter} {14}\BibitemShut {NoStop}%
\bibitem [{\citenamefont {Green}(2010)}]{Green:2010}%
  \BibitemOpen
  \bibfield  {author} {\bibinfo {author} {\bibfnamefont {A.~M.}\ \bibnamefont
  {Green}},\ }\href {\doibase 10.1088/1475-7516/2010/10/034} {\bibfield
  {journal} {\bibinfo  {journal} {J. Cosmol. Astropart. Phys.}\ }\textbf
  {\bibinfo {volume} {10}},\ \bibinfo {pages} {034} (\bibinfo {year} {2010})},\
  \Eprint {http://arxiv.org/abs/1009.0916} {arXiv:1009.0916} \BibitemShut
  {NoStop}%
\bibitem [{\citenamefont {Peter}(2011)}]{Peter:2011}%
  \BibitemOpen
  \bibfield  {author} {\bibinfo {author} {\bibfnamefont {A.~H.~G.}\
  \bibnamefont {Peter}},\ }\href {http://dx.doi.org/10.1103/PhysRevD.83.125029}
  {\bibfield  {journal} {\bibinfo  {journal} {Phys. Rev. D}\ }\textbf {\bibinfo
  {volume} {83}},\ \bibinfo {pages} {125029} (\bibinfo {year} {2011})},\
  \Eprint {http://arxiv.org/abs/1103.5145} {arXiv:1103.5145} \BibitemShut
  {NoStop}%
\bibitem [{\citenamefont {Fairbairn}\ \emph {et~al.}(2013)\citenamefont
  {Fairbairn}, \citenamefont {Douce},\ and\ \citenamefont
  {Swift}}]{Fairbairn:2012}%
  \BibitemOpen
  \bibfield  {author} {\bibinfo {author} {\bibfnamefont {M.}~\bibnamefont
  {Fairbairn}}, \bibinfo {author} {\bibfnamefont {T.}~\bibnamefont {Douce}}, \
  and\ \bibinfo {author} {\bibfnamefont {J.}~\bibnamefont {Swift}},\ }\href
  {\doibase 10.1016/j.astropartphys.2013.06.003} {\bibfield  {journal}
  {\bibinfo  {journal} {J. Astropart. Phys.}\ }\textbf {\bibinfo {volume}
  {47}},\ \bibinfo {pages} {45} (\bibinfo {year} {2013})},\ \Eprint
  {http://arxiv.org/abs/1206.2693} {arXiv:1206.2693} \BibitemShut {NoStop}%
\bibitem [{\citenamefont {Lisanti}\ \emph {et~al.}(2010)\citenamefont
  {Lisanti}, \citenamefont {Strigari}, \citenamefont {Wacker},\ and\
  \citenamefont {Wechsler}}]{Lisanti:2010}%
  \BibitemOpen
  \bibfield  {author} {\bibinfo {author} {\bibfnamefont {M.}~\bibnamefont
  {Lisanti}}, \bibinfo {author} {\bibfnamefont {L.~E.}\ \bibnamefont
  {Strigari}}, \bibinfo {author} {\bibfnamefont {J.~G.}\ \bibnamefont
  {Wacker}}, \ and\ \bibinfo {author} {\bibfnamefont {R.~H.}\ \bibnamefont
  {Wechsler}},\ }\href {http://dx.doi.org/10.1103/PhysRevD.83.023519}
  {\bibfield  {journal} {\bibinfo  {journal} {Phys. Rev. D}\ }\textbf {\bibinfo
  {volume} {83}},\ \bibinfo {pages} {023519} (\bibinfo {year} {2010})},\
  \Eprint {http://arxiv.org/abs/1010.4300} {arXiv:1010.4300} \BibitemShut
  {NoStop}%
\bibitem [{\citenamefont {Bhattacharjee}\ \emph {et~al.}(2013)\citenamefont
  {Bhattacharjee}, \citenamefont {Chaudhury}, \citenamefont {Kundu},\ and\
  \citenamefont {Majumdar}}]{Bhattacharjee:2012}%
  \BibitemOpen
  \bibfield  {author} {\bibinfo {author} {\bibfnamefont {P.}~\bibnamefont
  {Bhattacharjee}}, \bibinfo {author} {\bibfnamefont {S.}~\bibnamefont
  {Chaudhury}}, \bibinfo {author} {\bibfnamefont {S.}~\bibnamefont {Kundu}}, \
  and\ \bibinfo {author} {\bibfnamefont {S.}~\bibnamefont {Majumdar}},\ }\href
  {http://dx.doi.org/10.1103/PhysRevD.87.083525} {\bibfield  {journal}
  {\bibinfo  {journal} {Phys. Rev. D}\ }\textbf {\bibinfo {volume} {87}},\
  \bibinfo {pages} {083525} (\bibinfo {year} {2013})},\ \Eprint
  {http://arxiv.org/abs/1210.2328} {arXiv:1210.2328} \BibitemShut {NoStop}%
\bibitem [{\citenamefont {Vogelsberger}\ \emph {et~al.}(2009)\citenamefont
  {Vogelsberger}, \citenamefont {Helmi}, \citenamefont {Springel},
  \citenamefont {White}, \citenamefont {Wang}, \citenamefont {Frenk},
  \citenamefont {Jenkins}, \citenamefont {Ludlow},\ and\ \citenamefont
  {Navarro}}]{Vogelsberger:2009}%
  \BibitemOpen
  \bibfield  {author} {\bibinfo {author} {\bibfnamefont {M.}~\bibnamefont
  {Vogelsberger}}, \bibinfo {author} {\bibfnamefont {A.}~\bibnamefont {Helmi}},
  \bibinfo {author} {\bibfnamefont {V.}~\bibnamefont {Springel}}, \bibinfo
  {author} {\bibfnamefont {S.~D.~M.}\ \bibnamefont {White}}, \bibinfo {author}
  {\bibfnamefont {J.}~\bibnamefont {Wang}}, \bibinfo {author} {\bibfnamefont
  {C.~S.}\ \bibnamefont {Frenk}}, \bibinfo {author} {\bibfnamefont
  {A.}~\bibnamefont {Jenkins}}, \bibinfo {author} {\bibfnamefont
  {A.}~\bibnamefont {Ludlow}}, \ and\ \bibinfo {author} {\bibfnamefont {J.~F.}\
  \bibnamefont {Navarro}},\ }\href {\doibase 10.1111/j.1365-2966.2009.14630.x}
  {\bibfield  {journal} {\bibinfo  {journal} {Mon. Not. R. Astron. Soc.}\
  }\textbf {\bibinfo {volume} {395}},\ \bibinfo {pages} {797} (\bibinfo {year}
  {2009})},\ \Eprint {http://arxiv.org/abs/0812.0362} {arXiv:0812.0362}
  \BibitemShut {NoStop}%
\bibitem [{\citenamefont {Kuhlen}\ \emph {et~al.}(2010)\citenamefont {Kuhlen},
  \citenamefont {Weiner}, \citenamefont {Diemand}, \citenamefont {Madau},
  \citenamefont {Moore}, \citenamefont {Potter}, \citenamefont {Stadel},\ and\
  \citenamefont {Zemp}}]{Kuhlen:2010}%
  \BibitemOpen
  \bibfield  {author} {\bibinfo {author} {\bibfnamefont {M.}~\bibnamefont
  {Kuhlen}}, \bibinfo {author} {\bibfnamefont {N.}~\bibnamefont {Weiner}},
  \bibinfo {author} {\bibfnamefont {J.}~\bibnamefont {Diemand}}, \bibinfo
  {author} {\bibfnamefont {P.}~\bibnamefont {Madau}}, \bibinfo {author}
  {\bibfnamefont {B.}~\bibnamefont {Moore}}, \bibinfo {author} {\bibfnamefont
  {D.}~\bibnamefont {Potter}}, \bibinfo {author} {\bibfnamefont
  {J.}~\bibnamefont {Stadel}}, \ and\ \bibinfo {author} {\bibfnamefont
  {M.}~\bibnamefont {Zemp}},\ }\href {\doibase 10.1088/1475-7516/2010/02/030}
  {\bibfield  {journal} {\bibinfo  {journal} {J. Cosmol. Astropart. Phys.}\
  }\textbf {\bibinfo {volume} {02}},\ \bibinfo {pages} {030} (\bibinfo {year}
  {2010})},\ \Eprint {http://arxiv.org/abs/0912.2358} {arXiv:0912.2358}
  \BibitemShut {NoStop}%
\bibitem [{\citenamefont {Kuhlen}\ \emph {et~al.}(2012)\citenamefont {Kuhlen},
  \citenamefont {Lisanti},\ and\ \citenamefont {Spergel}}]{Kuhlen:2012}%
  \BibitemOpen
  \bibfield  {author} {\bibinfo {author} {\bibfnamefont {M.}~\bibnamefont
  {Kuhlen}}, \bibinfo {author} {\bibfnamefont {M.}~\bibnamefont {Lisanti}}, \
  and\ \bibinfo {author} {\bibfnamefont {D.~N.}\ \bibnamefont {Spergel}},\
  }\href {\doibase 10.1103/PhysRevD.86.063505} {\bibfield  {journal} {\bibinfo
  {journal} {Phys. Rev. D}\ }\textbf {\bibinfo {volume} {86}},\ \bibinfo
  {pages} {063505} (\bibinfo {year} {2012})},\ \Eprint
  {http://arxiv.org/abs/1202.0007} {arXiv:1202.0007} \BibitemShut {NoStop}%
\bibitem [{\citenamefont {Mao}\ \emph {et~al.}(2012)\citenamefont {Mao},
  \citenamefont {Strigari}, \citenamefont {Wechsler}, \citenamefont {Wu},\ and\
  \citenamefont {Hahn}}]{Mao:2012}%
  \BibitemOpen
  \bibfield  {author} {\bibinfo {author} {\bibfnamefont {Y.-Y.}\ \bibnamefont
  {Mao}}, \bibinfo {author} {\bibfnamefont {L.~E.}\ \bibnamefont {Strigari}},
  \bibinfo {author} {\bibfnamefont {R.~H.}\ \bibnamefont {Wechsler}}, \bibinfo
  {author} {\bibfnamefont {H.-Y.}\ \bibnamefont {Wu}}, \ and\ \bibinfo {author}
  {\bibfnamefont {O.}~\bibnamefont {Hahn}},\ }\href {\doibase
  10.1088/0004-637x/764/1/35} {\bibfield  {journal} {\bibinfo  {journal}
  {Astrophys. J.}\ }\textbf {\bibinfo {volume} {764}},\ \bibinfo {pages} {35}
  (\bibinfo {year} {2012})},\ \Eprint {http://arxiv.org/abs/1210.2721}
  {arXiv:1210.2721} \BibitemShut {NoStop}%
\bibitem [{\citenamefont {Read}\ \emph {et~al.}(2009)\citenamefont {Read},
  \citenamefont {Mayer}, \citenamefont {Brooks}, \citenamefont {Governato},\
  and\ \citenamefont {Lake}}]{Read:2009}%
  \BibitemOpen
  \bibfield  {author} {\bibinfo {author} {\bibfnamefont {J.~I.}\ \bibnamefont
  {Read}}, \bibinfo {author} {\bibfnamefont {L.}~\bibnamefont {Mayer}},
  \bibinfo {author} {\bibfnamefont {A.~M.}\ \bibnamefont {Brooks}}, \bibinfo
  {author} {\bibfnamefont {F.}~\bibnamefont {Governato}}, \ and\ \bibinfo
  {author} {\bibfnamefont {G.}~\bibnamefont {Lake}},\ }\href {\doibase
  10.1111/j.1365-2966.2009.14757.x} {\bibfield  {journal} {\bibinfo  {journal}
  {Mon. Not. R. Astron. Soc.}\ }\textbf {\bibinfo {volume} {397}},\ \bibinfo
  {pages} {44} (\bibinfo {year} {2009})},\ \Eprint
  {http://arxiv.org/abs/0902.0009} {arXiv:0902.0009} \BibitemShut {NoStop}%
\bibitem [{\citenamefont {Read}\ \emph {et~al.}(2010)\citenamefont {Read} \emph
  {et~al.}}]{Read:2010}%
  \BibitemOpen
  \bibfield  {author} {\bibinfo {author} {\bibfnamefont {J.~I.}\ \bibnamefont
  {Read}} \emph {et~al.},\ }\href {\doibase 10.1063/1.3458542} {\bibfield
  {journal} {\bibinfo  {journal} {AIP Conf. Proc.}\ }\textbf {\bibinfo {volume}
  {1240}},\ \bibinfo {pages} {391} (\bibinfo {year} {2010})},\ \Eprint
  {http://arxiv.org/abs/0901.2938} {arXiv:0901.2938} \BibitemShut {NoStop}%
\bibitem [{\citenamefont {Kuhlen}\ \emph {et~al.}(2013)\citenamefont {Kuhlen},
  \citenamefont {Pillepich}, \citenamefont {Guedes},\ and\ \citenamefont
  {Madau}}]{Kuhlen:2013}%
  \BibitemOpen
  \bibfield  {author} {\bibinfo {author} {\bibfnamefont {M.}~\bibnamefont
  {Kuhlen}}, \bibinfo {author} {\bibfnamefont {A.}~\bibnamefont {Pillepich}},
  \bibinfo {author} {\bibfnamefont {J.}~\bibnamefont {Guedes}}, \ and\ \bibinfo
  {author} {\bibfnamefont {P.}~\bibnamefont {Madau}},\ }\href
  {http://arxiv.org/abs/1308.1703} {\  (\bibinfo {year} {2013})},\ \Eprint
  {http://arxiv.org/abs/1308.1703} {arXiv:1308.1703} \BibitemShut {NoStop}%
\bibitem [{\citenamefont {Strigari}\ and\ \citenamefont
  {Trotta}(2009)}]{Strigari:2009}%
  \BibitemOpen
  \bibfield  {author} {\bibinfo {author} {\bibfnamefont {L.~E.}\ \bibnamefont
  {Strigari}}\ and\ \bibinfo {author} {\bibfnamefont {R.}~\bibnamefont
  {Trotta}},\ }\href {\doibase 10.1088/1475-7516/2009/11/019} {\bibfield
  {journal} {\bibinfo  {journal} {J. Cosmol. Astropart. Phys.}\ }\textbf
  {\bibinfo {volume} {11}},\ \bibinfo {pages} {019} (\bibinfo {year} {2009})},\
  \Eprint {http://arxiv.org/abs/0906.5361} {arXiv:0906.5361} \BibitemShut
  {NoStop}%
\bibitem [{\citenamefont {Peter}(2010)}]{Peter:2009}%
  \BibitemOpen
  \bibfield  {author} {\bibinfo {author} {\bibfnamefont {A.~H.~G.}\
  \bibnamefont {Peter}},\ }\href {\doibase 10.1103/PhysRevD.81.087301}
  {\bibfield  {journal} {\bibinfo  {journal} {Phys. Rev. D}\ }\textbf {\bibinfo
  {volume} {81}},\ \bibinfo {pages} {087301} (\bibinfo {year} {2010})},\
  \Eprint {http://arxiv.org/abs/0910.4765} {arXiv:0910.4765} \BibitemShut
  {NoStop}%
\bibitem [{\citenamefont {Pato}\ \emph {et~al.}(2011)\citenamefont {Pato},
  \citenamefont {Baudis}, \citenamefont {Bertone}, \citenamefont {de~Austri},
  \citenamefont {Strigari},\ and\ \citenamefont {Trotta}}]{Pato:2011}%
  \BibitemOpen
  \bibfield  {author} {\bibinfo {author} {\bibfnamefont {M.}~\bibnamefont
  {Pato}}, \bibinfo {author} {\bibfnamefont {L.}~\bibnamefont {Baudis}},
  \bibinfo {author} {\bibfnamefont {G.}~\bibnamefont {Bertone}}, \bibinfo
  {author} {\bibfnamefont {R.~R.}\ \bibnamefont {de~Austri}}, \bibinfo {author}
  {\bibfnamefont {L.~E.}\ \bibnamefont {Strigari}}, \ and\ \bibinfo {author}
  {\bibfnamefont {R.}~\bibnamefont {Trotta}},\ }\href
  {http://dx.doi.org/10.1103/physrevd.83.083505} {\bibfield  {journal}
  {\bibinfo  {journal} {Phys. Rev. D}\ }\textbf {\bibinfo {volume} {83}},\
  \bibinfo {pages} {083505} (\bibinfo {year} {2011})},\ \Eprint
  {http://arxiv.org/abs/1012.3458} {arXiv:1012.3458} \BibitemShut {NoStop}%
\bibitem [{\citenamefont {Pato}\ \emph {et~al.}(2013)\citenamefont {Pato},
  \citenamefont {Strigari}, \citenamefont {Trotta},\ and\ \citenamefont
  {Bertone}}]{Pato:2013}%
  \BibitemOpen
  \bibfield  {author} {\bibinfo {author} {\bibfnamefont {M.}~\bibnamefont
  {Pato}}, \bibinfo {author} {\bibfnamefont {L.~E.}\ \bibnamefont {Strigari}},
  \bibinfo {author} {\bibfnamefont {R.}~\bibnamefont {Trotta}}, \ and\ \bibinfo
  {author} {\bibfnamefont {G.}~\bibnamefont {Bertone}},\ }\href {\doibase
  10.1088/1475-7516/2013/02/041} {\bibfield  {journal} {\bibinfo  {journal} {J.
  Cosmol. Astropart. Phys.}\ }\textbf {\bibinfo {volume} {02}},\ \bibinfo
  {pages} {041} (\bibinfo {year} {2013})},\ \Eprint
  {http://arxiv.org/abs/1211.7063} {arXiv:1211.7063} \BibitemShut {NoStop}%
\bibitem [{\citenamefont {Drees}\ and\ \citenamefont
  {Shan}(2007)}]{Drees:2007}%
  \BibitemOpen
  \bibfield  {author} {\bibinfo {author} {\bibfnamefont {M.}~\bibnamefont
  {Drees}}\ and\ \bibinfo {author} {\bibfnamefont {C.-L.}\ \bibnamefont
  {Shan}},\ }\href {\doibase 10.1088/1475-7516/2007/06/011} {\bibfield
  {journal} {\bibinfo  {journal} {J. Cosmol. Astropart. Phys.}\ }\textbf
  {\bibinfo {volume} {06}},\ \bibinfo {pages} {011} (\bibinfo {year} {2007})},\
  \Eprint {http://arxiv.org/abs/astro-ph/0703651} {astro-ph/0703651}
  \BibitemShut {NoStop}%
\bibitem [{\citenamefont {Drees}\ and\ \citenamefont
  {Shan}(2008)}]{Drees:2008}%
  \BibitemOpen
  \bibfield  {author} {\bibinfo {author} {\bibfnamefont {M.}~\bibnamefont
  {Drees}}\ and\ \bibinfo {author} {\bibfnamefont {C.-L.}\ \bibnamefont
  {Shan}},\ }\href {\doibase 10.1088/1475-7516/2008/06/012} {\bibfield
  {journal} {\bibinfo  {journal} {J. Cosmol. Astropart. Phys.}\ }\textbf
  {\bibinfo {volume} {06}},\ \bibinfo {pages} {012} (\bibinfo {year} {2008})},\
  \Eprint {http://arxiv.org/abs/0803.4477} {arXiv:0803.4477} \BibitemShut
  {NoStop}%
\bibitem [{\citenamefont {Kavanagh}\ and\ \citenamefont
  {Green}(2013)}]{Kavanagh:2013a}%
  \BibitemOpen
  \bibfield  {author} {\bibinfo {author} {\bibfnamefont {B.~J.}\ \bibnamefont
  {Kavanagh}}\ and\ \bibinfo {author} {\bibfnamefont {A.~M.}\ \bibnamefont
  {Green}},\ }\href {http://dx.doi.org/10.1103/physrevlett.111.031302}
  {\bibfield  {journal} {\bibinfo  {journal} {Phys. Rev. Lett.}\ }\textbf
  {\bibinfo {volume} {111}},\ \bibinfo {pages} {031302} (\bibinfo {year}
  {2013})},\ \Eprint {http://arxiv.org/abs/1303.6868} {arXiv:1303.6868}
  \BibitemShut {NoStop}%
\bibitem [{\citenamefont {Lewin}\ and\ \citenamefont
  {Smith}(1996)}]{Lewin:1996}%
  \BibitemOpen
  \bibfield  {author} {\bibinfo {author} {\bibfnamefont {J.~D.}\ \bibnamefont
  {Lewin}}\ and\ \bibinfo {author} {\bibfnamefont {P.~F.}\ \bibnamefont
  {Smith}},\ }\href {\doibase 10.1016/s0927-6505(96)00047-3} {\bibfield
  {journal} {\bibinfo  {journal} {J. Astropart. Phys.}\ }\textbf {\bibinfo
  {volume} {6}},\ \bibinfo {pages} {87} (\bibinfo {year} {1996})}\BibitemShut
  {NoStop}%
\bibitem [{\citenamefont {Helm}(1956)}]{Helm:1956}%
  \BibitemOpen
  \bibfield  {author} {\bibinfo {author} {\bibfnamefont {R.}~\bibnamefont
  {Helm}},\ }\href {http://dx.doi.org/10.1103/PhysRev.104.1466} {\bibfield
  {journal} {\bibinfo  {journal} {Phys. Rev.}\ }\textbf {\bibinfo {volume}
  {104}},\ \bibinfo {pages} {1466} (\bibinfo {year} {1956})}\BibitemShut
  {NoStop}%
\bibitem [{\citenamefont {Smith}\ and\ \citenamefont
  {Weiner}(2001)}]{Smith:2001}%
  \BibitemOpen
  \bibfield  {author} {\bibinfo {author} {\bibfnamefont {D.}~\bibnamefont
  {Smith}}\ and\ \bibinfo {author} {\bibfnamefont {N.}~\bibnamefont {Weiner}},\
  }\href {http://dx.doi.org/10.1103/physrevd.64.043502} {\bibfield  {journal}
  {\bibinfo  {journal} {Phys. Rev. D}\ }\textbf {\bibinfo {volume} {64}},\
  \bibinfo {pages} {043502} (\bibinfo {year} {2001})},\ \Eprint
  {http://arxiv.org/abs/hep-ph/0101138} {hep-ph/0101138} \BibitemShut {NoStop}%
\bibitem [{\citenamefont {Kurylov}\ and\ \citenamefont
  {Kamionkowski}(2003)}]{Kurylov:2003}%
  \BibitemOpen
  \bibfield  {author} {\bibinfo {author} {\bibfnamefont {A.}~\bibnamefont
  {Kurylov}}\ and\ \bibinfo {author} {\bibfnamefont {M.}~\bibnamefont
  {Kamionkowski}},\ }\href {http://dx.doi.org/10.1103/physrevd.69.063503}
  {\bibfield  {journal} {\bibinfo  {journal} {Phys. Rev. D}\ }\textbf {\bibinfo
  {volume} {69}},\ \bibinfo {pages} {063503} (\bibinfo {year} {2003})},\
  \Eprint {http://arxiv.org/abs/hep-ph/0307185} {hep-ph/0307185} \BibitemShut
  {NoStop}%
\bibitem [{\citenamefont {Fan}\ \emph {et~al.}(2010)\citenamefont {Fan},
  \citenamefont {Reece},\ and\ \citenamefont {Wang}}]{Fan:2010}%
  \BibitemOpen
  \bibfield  {author} {\bibinfo {author} {\bibfnamefont {J.}~\bibnamefont
  {Fan}}, \bibinfo {author} {\bibfnamefont {M.}~\bibnamefont {Reece}}, \ and\
  \bibinfo {author} {\bibfnamefont {L.-T.}\ \bibnamefont {Wang}},\ }\href
  {\doibase 10.1088/1475-7516/2010/11/042} {\bibfield  {journal} {\bibinfo
  {journal} {J. Cosmol. Astropart. Phys.}\ }\textbf {\bibinfo {volume} {11}},\
  \bibinfo {pages} {042} (\bibinfo {year} {2010})},\ \Eprint
  {http://arxiv.org/abs/1008.1591} {arXiv:1008.1591} \BibitemShut {NoStop}%
\bibitem [{\citenamefont {Fitzpatrick}\ \emph {et~al.}(2012)\citenamefont
  {Fitzpatrick}, \citenamefont {Haxton}, \citenamefont {Katz}, \citenamefont
  {Lubbers},\ and\ \citenamefont {Xu}}]{Fitzpatrick:2012}%
  \BibitemOpen
  \bibfield  {author} {\bibinfo {author} {\bibfnamefont {A.~L.}\ \bibnamefont
  {Fitzpatrick}}, \bibinfo {author} {\bibfnamefont {W.}~\bibnamefont {Haxton}},
  \bibinfo {author} {\bibfnamefont {E.}~\bibnamefont {Katz}}, \bibinfo {author}
  {\bibfnamefont {N.}~\bibnamefont {Lubbers}}, \ and\ \bibinfo {author}
  {\bibfnamefont {Y.}~\bibnamefont {Xu}},\ }\href
  {http://arxiv.org/abs/1211.2818} {\  (\bibinfo {year} {2012})},\ \Eprint
  {http://arxiv.org/abs/1211.2818} {arXiv:1211.2818} \BibitemShut {NoStop}%
\bibitem [{\citenamefont {Fitzpatrick}\ \emph {et~al.}(2013)\citenamefont
  {Fitzpatrick}, \citenamefont {Haxton}, \citenamefont {Katz}, \citenamefont
  {Lubbers},\ and\ \citenamefont {Xu}}]{Fitzpatrick:2013}%
  \BibitemOpen
  \bibfield  {author} {\bibinfo {author} {\bibfnamefont {A.~L.}\ \bibnamefont
  {Fitzpatrick}}, \bibinfo {author} {\bibfnamefont {W.}~\bibnamefont {Haxton}},
  \bibinfo {author} {\bibfnamefont {E.}~\bibnamefont {Katz}}, \bibinfo {author}
  {\bibfnamefont {N.}~\bibnamefont {Lubbers}}, \ and\ \bibinfo {author}
  {\bibfnamefont {Y.}~\bibnamefont {Xu}},\ }\href {\doibase
  10.1088/1475-7516/2013/02/004} {\bibfield  {journal} {\bibinfo  {journal} {J.
  Cosmol. Astropart. Phys.}\ }\textbf {\bibinfo {volume} {02}},\ \bibinfo
  {pages} {004} (\bibinfo {year} {2013})},\ \Eprint
  {http://arxiv.org/abs/1203.3542} {arXiv:1203.3542} \BibitemShut {NoStop}%
\bibitem [{\citenamefont {Freese}\ \emph {et~al.}(2013)\citenamefont {Freese},
  \citenamefont {Lisanti},\ and\ \citenamefont {Savage}}]{Freese:2013}%
  \BibitemOpen
  \bibfield  {author} {\bibinfo {author} {\bibfnamefont {K.}~\bibnamefont
  {Freese}}, \bibinfo {author} {\bibfnamefont {M.}~\bibnamefont {Lisanti}}, \
  and\ \bibinfo {author} {\bibfnamefont {C.}~\bibnamefont {Savage}},\ }\href
  {http://arxiv.org/abs/1209.3339} {\  (\bibinfo {year} {2013})},\ \Eprint
  {http://arxiv.org/abs/1209.3339} {arXiv:1209.3339} \BibitemShut {NoStop}%
\bibitem [{\citenamefont {Smith}\ \emph {et~al.}(2007)\citenamefont {Smith}
  \emph {et~al.}}]{RAVE:2007}%
  \BibitemOpen
  \bibfield  {author} {\bibinfo {author} {\bibfnamefont {M.~C.}\ \bibnamefont
  {Smith}} \emph {et~al.},\ }\href {\doibase 10.1111/j.1365-2966.2007.11964.x}
  {\bibfield  {journal} {\bibinfo  {journal} {Mon. Not. R. Astron. Soc.}\
  }\textbf {\bibinfo {volume} {379}},\ \bibinfo {pages} {755} (\bibinfo {year}
  {2007})},\ \Eprint {http://arxiv.org/abs/astro-ph/0611671} {astro-ph/0611671}
  \BibitemShut {NoStop}%
\bibitem [{\citenamefont {Piffl}\ \emph {et~al.}(2013)\citenamefont {Piffl}
  \emph {et~al.}}]{RAVE:2013}%
  \BibitemOpen
  \bibfield  {author} {\bibinfo {author} {\bibfnamefont {T.}~\bibnamefont
  {Piffl}} \emph {et~al.},\ }\href {http://arxiv.org/abs/1309.4293} {\
  (\bibinfo {year} {2013})},\ \Eprint {http://arxiv.org/abs/1309.4293}
  {arXiv:1309.4293} \BibitemShut {NoStop}%
\bibitem [{\citenamefont {Feroz}\ \emph {et~al.}(2008)\citenamefont {Feroz},
  \citenamefont {Hobson},\ and\ \citenamefont {Bridges}}]{MultiNest1}%
  \BibitemOpen
  \bibfield  {author} {\bibinfo {author} {\bibfnamefont {F.}~\bibnamefont
  {Feroz}}, \bibinfo {author} {\bibfnamefont {M.~P.}\ \bibnamefont {Hobson}}, \
  and\ \bibinfo {author} {\bibfnamefont {M.}~\bibnamefont {Bridges}},\ }\href
  {\doibase 10.1111/j.1365-2966.2009.14548.x} {\bibfield  {journal} {\bibinfo
  {journal} {Mon. Not. R. Astron. Soc.}\ }\textbf {\bibinfo {volume} {398}},\
  \bibinfo {pages} {1601} (\bibinfo {year} {2008})},\ \Eprint
  {http://arxiv.org/abs/0809.3437} {arXiv:0809.3437} \BibitemShut {NoStop}%
\bibitem [{\citenamefont {Feroz}\ and\ \citenamefont
  {Hobson}(2007)}]{MultiNest2}%
  \BibitemOpen
  \bibfield  {author} {\bibinfo {author} {\bibfnamefont {F.}~\bibnamefont
  {Feroz}}\ and\ \bibinfo {author} {\bibfnamefont {M.~P.}\ \bibnamefont
  {Hobson}},\ }\href {\doibase 10.1111/j.1365-2966.2007.12353.x} {\bibfield
  {journal} {\bibinfo  {journal} {Mon. Not. R. Astron. Soc.}\ }\textbf
  {\bibinfo {volume} {384}},\ \bibinfo {pages} {449} (\bibinfo {year}
  {2007})},\ \Eprint {http://arxiv.org/abs/0704.3704} {arXiv:0704.3704}
  \BibitemShut {NoStop}%
\bibitem [{\citenamefont {Feroz}\ \emph {et~al.}(2013)\citenamefont {Feroz},
  \citenamefont {Hobson}, \citenamefont {Cameron},\ and\ \citenamefont
  {Pettitt}}]{MultiNest3}%
  \BibitemOpen
  \bibfield  {author} {\bibinfo {author} {\bibfnamefont {F.}~\bibnamefont
  {Feroz}}, \bibinfo {author} {\bibfnamefont {M.~P.}\ \bibnamefont {Hobson}},
  \bibinfo {author} {\bibfnamefont {E.}~\bibnamefont {Cameron}}, \ and\
  \bibinfo {author} {\bibfnamefont {A.~N.}\ \bibnamefont {Pettitt}},\ }\href
  {http://arxiv.org/abs/1306.2144} {\  (\bibinfo {year} {2013})},\ \Eprint
  {http://arxiv.org/abs/1306.2144} {arXiv:1306.2144} \BibitemShut {NoStop}%
\bibitem [{\citenamefont {Kavanagh}\ and\ \citenamefont
  {Green}(2012)}]{Kavanagh:2012}%
  \BibitemOpen
  \bibfield  {author} {\bibinfo {author} {\bibfnamefont {B.~J.}\ \bibnamefont
  {Kavanagh}}\ and\ \bibinfo {author} {\bibfnamefont {A.~M.}\ \bibnamefont
  {Green}},\ }\href {http://dx.doi.org/10.1103/PhysRevD.86.065027} {\bibfield
  {journal} {\bibinfo  {journal} {Phys. Rev. D}\ }\textbf {\bibinfo {volume}
  {86}},\ \bibinfo {pages} {065027} (\bibinfo {year} {2012})},\ \Eprint
  {http://arxiv.org/abs/1207.2039} {arXiv:1207.2039} \BibitemShut {NoStop}%
\bibitem [{\citenamefont {Iocco}\ \emph {et~al.}(2011)\citenamefont {Iocco},
  \citenamefont {Pato}, \citenamefont {Bertone},\ and\ \citenamefont
  {Jetzer}}]{Iocco:2011}%
  \BibitemOpen
  \bibfield  {author} {\bibinfo {author} {\bibfnamefont {F.}~\bibnamefont
  {Iocco}}, \bibinfo {author} {\bibfnamefont {M.}~\bibnamefont {Pato}},
  \bibinfo {author} {\bibfnamefont {G.}~\bibnamefont {Bertone}}, \ and\
  \bibinfo {author} {\bibfnamefont {P.}~\bibnamefont {Jetzer}},\ }\href
  {\doibase 10.1088/1475-7516/2011/11/029} {\bibfield  {journal} {\bibinfo
  {journal} {J. Cosmol. Astropart. Phys.}\ }\textbf {\bibinfo {volume} {11}},\
  \bibinfo {pages} {029} (\bibinfo {year} {2011})},\ \Eprint
  {http://arxiv.org/abs/1107.5810} {arXiv:1107.5810} \BibitemShut {NoStop}%
\bibitem [{\citenamefont {Bovy}\ and\ \citenamefont
  {Tremaine}(2012)}]{Bovy:2012}%
  \BibitemOpen
  \bibfield  {author} {\bibinfo {author} {\bibfnamefont {J.}~\bibnamefont
  {Bovy}}\ and\ \bibinfo {author} {\bibfnamefont {S.}~\bibnamefont
  {Tremaine}},\ }\href {\doibase 10.1088/0004-637x/756/1/89} {\bibfield
  {journal} {\bibinfo  {journal} {Astrophys. J.}\ }\textbf {\bibinfo {volume}
  {756}},\ \bibinfo {pages} {89} (\bibinfo {year} {2012})},\ \Eprint
  {http://arxiv.org/abs/1205.4033} {arXiv:1205.4033} \BibitemShut {NoStop}%
\bibitem [{\citenamefont {Zhang}\ \emph {et~al.}(2013)\citenamefont {Zhang},
  \citenamefont {Rix}, \citenamefont {van~de Ven}, \citenamefont {Bovy},
  \citenamefont {Liu},\ and\ \citenamefont {Zhao}}]{Zhang:2013}%
  \BibitemOpen
  \bibfield  {author} {\bibinfo {author} {\bibfnamefont {L.}~\bibnamefont
  {Zhang}}, \bibinfo {author} {\bibfnamefont {H.-W.}\ \bibnamefont {Rix}},
  \bibinfo {author} {\bibfnamefont {G.}~\bibnamefont {van~de Ven}}, \bibinfo
  {author} {\bibfnamefont {J.}~\bibnamefont {Bovy}}, \bibinfo {author}
  {\bibfnamefont {C.}~\bibnamefont {Liu}}, \ and\ \bibinfo {author}
  {\bibfnamefont {G.}~\bibnamefont {Zhao}},\ }\href {\doibase
  10.1088/0004-637X/772/2/108} {\bibfield  {journal} {\bibinfo  {journal}
  {Astrophys. J.}\ }\textbf {\bibinfo {volume} {772}},\ \bibinfo {pages} {108}
  (\bibinfo {year} {2013})},\ \Eprint {http://arxiv.org/abs/1209.0256}
  {arXiv:1209.0256} \BibitemShut {NoStop}%
\bibitem [{\citenamefont {Nesti}\ and\ \citenamefont
  {Salucci}(2013)}]{Nesti:2013}%
  \BibitemOpen
  \bibfield  {author} {\bibinfo {author} {\bibfnamefont {F.}~\bibnamefont
  {Nesti}}\ and\ \bibinfo {author} {\bibfnamefont {P.}~\bibnamefont
  {Salucci}},\ }\href {\doibase 10.1088/1475-7516/2013/07/016} {\bibfield
  {journal} {\bibinfo  {journal} {J. Cosmol. Astropart. Phys.}\ }\textbf
  {\bibinfo {volume} {07}},\ \bibinfo {pages} {016} (\bibinfo {year} {2013})},\
  \Eprint {http://arxiv.org/abs/1304.5127} {arXiv:1304.5127} \BibitemShut
  {NoStop}%
\bibitem [{\citenamefont {Peter}\ \emph {et~al.}(2013)\citenamefont {Peter},
  \citenamefont {Gluscevic}, \citenamefont {Green}, \citenamefont {Kavanagh},\
  and\ \citenamefont {Lee}}]{Peter:2013a}%
  \BibitemOpen
  \bibfield  {author} {\bibinfo {author} {\bibfnamefont {A.~H.~G.}\
  \bibnamefont {Peter}}, \bibinfo {author} {\bibfnamefont {V.}~\bibnamefont
  {Gluscevic}}, \bibinfo {author} {\bibfnamefont {A.~M.}\ \bibnamefont
  {Green}}, \bibinfo {author} {\bibfnamefont {B.~J.}\ \bibnamefont {Kavanagh}},
  \ and\ \bibinfo {author} {\bibfnamefont {S.~K.}\ \bibnamefont {Lee}},\ }\href
  {http://arxiv.org/abs/1310.7039} {\  (\bibinfo {year} {2013})},\ \Eprint
  {http://arxiv.org/abs/1310.7039} {arXiv:1310.7039} \BibitemShut {NoStop}%
\bibitem [{\citenamefont {Aprile}(2012)}]{Aprile:2012a}%
  \BibitemOpen
  \bibfield  {author} {\bibinfo {author} {\bibfnamefont {E.}~\bibnamefont
  {Aprile}} (\bibinfo {collaboration} {{XENON Collaboration}}),\ }\href
  {http://arxiv.org/abs/1206.6288} {\  (\bibinfo {year} {2012})},\ \Eprint
  {http://arxiv.org/abs/1206.6288} {arXiv:1206.6288} \BibitemShut {NoStop}%
\bibitem [{\citenamefont {{E. Aprile \textit{et al}.}}(2012)}]{Aprile:2012b}%
  \BibitemOpen
  \bibfield  {author} {\bibinfo {author} {\bibnamefont {{E. Aprile \textit{et
  al}.}}} (\bibinfo {collaboration} {{XENON100 Collaboration}}),\ }\href
  {\doibase 10.1103/PhysRevLett.109.181301} {\bibfield  {journal} {\bibinfo
  {journal} {Phys. Rev. Lett.}\ }\textbf {\bibinfo {volume} {109}},\ \bibinfo
  {pages} {181301} (\bibinfo {year} {2012})},\ \bibinfo {note} {online
  supplementary material},\ \Eprint {http://arxiv.org/abs/1207.5988}
  {arXiv:1207.5988} \BibitemShut {NoStop}%
\bibitem [{\citenamefont {Aprile}(2010)}]{Aprile:2010}%
  \BibitemOpen
  \bibfield  {author} {\bibinfo {author} {\bibfnamefont {E.}~\bibnamefont
  {Aprile}},\ }\href {www.bo.infn.it/xenon/docs/xe1t\_proposal\_v2.pdf}
  {\enquote {\bibinfo {title} {{XENON1T at LNGS: Proposal April 2010}},}\
  }\bibinfo {howpublished}
  {\url{www.bo.infn.it/xenon/docs/xe1t\_proposal\_v2.pdf}} (\bibinfo {year}
  {2010})\BibitemShut {NoStop}%
\bibitem [{\citenamefont {{P. Benetti \textit{et al.}}}(2007)}]{Benetti:2007}%
  \BibitemOpen
  \bibfield  {author} {\bibinfo {author} {\bibnamefont {{P. Benetti \textit{et
  al.}}}},\ }\href {\doibase 10.1016/j.astropartphys.2007.08.002} {\bibfield
  {journal} {\bibinfo  {journal} {J. Astropart. Phys.}\ }\textbf {\bibinfo
  {volume} {28}},\ \bibinfo {pages} {495} (\bibinfo {year} {2007})},\ \Eprint
  {http://arxiv.org/abs/astro-ph/0701286} {arXiv:astro-ph/0701286} \BibitemShut
  {NoStop}%
\bibitem [{\citenamefont {Grandi}(2005)}]{Grandi:2005}%
  \BibitemOpen
  \bibfield  {author} {\bibinfo {author} {\bibfnamefont {L.}~\bibnamefont
  {Grandi}},\ }\emph {\bibinfo {title} {{WARP: an argon double phase technique
  for Dark Matter search}}},\ \href@noop {} {Ph.D. thesis},\ \bibinfo  {school}
  {Department of Nuclear and Theoretical Physics, University of Pavia}
  (\bibinfo {year} {2005})\BibitemShut {NoStop}%
\bibitem [{\citenamefont {Bauer}(2013{\natexlab{a}})}]{Bauer:2013b}%
  \BibitemOpen
  \bibfield  {author} {\bibinfo {author} {\bibfnamefont {D.}~\bibnamefont
  {Bauer}},\ }\href@noop {} {\enquote {\bibinfo {title} {{SuperCDMS SNOLAB}},}\
  }\bibinfo {howpublished}
  {\url{https://indico.fnal.gov/getFile.py/access?sessionId=1\&resId=0\&materi%
alId=0\&confId=6584}} (\bibinfo {year} {2013}{\natexlab{a}})\BibitemShut
  {NoStop}%
\bibitem [{\citenamefont {Bauer}(2013{\natexlab{b}})}]{Bauer:2013a}%
  \BibitemOpen
  \bibfield  {author} {\bibinfo {author} {\bibfnamefont {D.}~\bibnamefont
  {Bauer}},\ }\href@noop {} {\enquote {\bibinfo {title} {{SuperCDMS update}},}\
  }\bibinfo {howpublished}
  {\url{www.fnal.gov/directorate/program\_planning/all\_experimenters\_meeting%
s/special\_reports/Bauer\_supercdms\_status\_02\_04\_13.pdf}} (\bibinfo {year}
  {2013}{\natexlab{b}})\BibitemShut {NoStop}%
\bibitem [{\citenamefont {Badertscher}\ \emph {et~al.}(2013)\citenamefont
  {Badertscher} \emph {et~al.}}]{Badertscher:2013}%
  \BibitemOpen
  \bibfield  {author} {\bibinfo {author} {\bibfnamefont {A.}~\bibnamefont
  {Badertscher}} \emph {et~al.} (\bibinfo {collaboration} {{ArDM
  Collaboration}}),\ }\href {http://arxiv.org/abs/1307.0117} {\  (\bibinfo
  {year} {2013})},\ \Eprint {http://arxiv.org/abs/1307.0117} {arXiv:1307.0117}
  \BibitemShut {NoStop}%
\bibitem [{\citenamefont {Cowan}\ \emph {et~al.}(2013)\citenamefont {Cowan},
  \citenamefont {Cranmer}, \citenamefont {Gross},\ and\ \citenamefont
  {Vitells}}]{Cowan:2013}%
  \BibitemOpen
  \bibfield  {author} {\bibinfo {author} {\bibfnamefont {G.}~\bibnamefont
  {Cowan}}, \bibinfo {author} {\bibfnamefont {K.}~\bibnamefont {Cranmer}},
  \bibinfo {author} {\bibfnamefont {E.}~\bibnamefont {Gross}}, \ and\ \bibinfo
  {author} {\bibfnamefont {O.}~\bibnamefont {Vitells}},\ }\href {\doibase
  10.1140/epjc/s10052-011-1554-0} {\bibfield  {journal} {\bibinfo  {journal}
  {Eur. Phys. J. C}\ }\textbf {\bibinfo {volume} {71}},\ \bibinfo {pages} {1}
  (\bibinfo {year} {2013})},\ \Eprint {http://arxiv.org/abs/1007.1727}
  {arXiv:1007.1727} \BibitemShut {NoStop}%
\bibitem [{\citenamefont {Aprile}\ \emph {et~al.}(2011)\citenamefont {Aprile}
  \emph {et~al.}}]{Aprile:2011}%
  \BibitemOpen
  \bibfield  {author} {\bibinfo {author} {\bibfnamefont {E.}~\bibnamefont
  {Aprile}} \emph {et~al.} (\bibinfo {collaboration} {{XENON100
  Collaboration}}),\ }\href {http://dx.doi.org/10.1103/PhysRevD.84.052003}
  {\bibfield  {journal} {\bibinfo  {journal} {Phys. Rev. D}\ }\textbf {\bibinfo
  {volume} {84}},\ \bibinfo {pages} {052003} (\bibinfo {year} {2011})},\
  \Eprint {http://arxiv.org/abs/1103.0303} {arXiv:1103.0303} \BibitemShut
  {NoStop}%
\bibitem [{\citenamefont {Ahmed}\ \emph {et~al.}(2009)\citenamefont {Ahmed}
  \emph {et~al.}}]{Ahmed:2009}%
  \BibitemOpen
  \bibfield  {author} {\bibinfo {author} {\bibfnamefont {Z.}~\bibnamefont
  {Ahmed}} \emph {et~al.} (\bibinfo {collaboration} {{CDMS Collaboration}}),\
  }\href {http://dx.doi.org/10.1103/PhysRevLett.102.011301} {\bibfield
  {journal} {\bibinfo  {journal} {Phys. Rev. Lett.}\ }\textbf {\bibinfo
  {volume} {102}},\ \bibinfo {pages} {011301} (\bibinfo {year} {2009})},\
  \Eprint {http://arxiv.org/abs/0802.3530} {arXiv:0802.3530} \BibitemShut
  {NoStop}%
\bibitem [{\citenamefont {Schwarz}(1978)}]{Schwarz:1978}%
  \BibitemOpen
  \bibfield  {author} {\bibinfo {author} {\bibfnamefont {G.}~\bibnamefont
  {Schwarz}},\ }\href {\doibase 10.1214/aos/1176344136} {\bibfield  {journal}
  {\bibinfo  {journal} {Ann. Stat.}\ }\textbf {\bibinfo {volume} {6}},\
  \bibinfo {pages} {461} (\bibinfo {year} {1978})}\BibitemShut {NoStop}%
\bibitem [{\citenamefont {Gautschi}(1978)}]{Gautschi:1978}%
  \BibitemOpen
  \bibfield  {author} {\bibinfo {author} {\bibfnamefont {W.}~\bibnamefont
  {Gautschi}},\ }in\ \href {http://www.worldcat.org/isbn/9780122083600} {\emph
  {\bibinfo {booktitle} {Recent advances in numerical analysis}}},\ \bibinfo
  {editor} {edited by\ \bibinfo {editor} {\bibfnamefont {C.}~\bibnamefont
  {De~Boor}}\ and\ \bibinfo {editor} {\bibfnamefont {G.~H.}\ \bibnamefont
  {Golub}}}\ (\bibinfo  {publisher} {Academic Press},\ \bibinfo {address} {New
  York},\ \bibinfo {year} {1978})\ pp.\ \bibinfo {pages} {45--72}\BibitemShut
  {NoStop}%
\bibitem [{\citenamefont {Wilkinson}(1984)}]{Wilkinson:1984}%
  \BibitemOpen
  \bibfield  {author} {\bibinfo {author} {\bibfnamefont {J.~H.}\ \bibnamefont
  {Wilkinson}},\ }in\ \href {http://www.worldcat.org/isbn/9780883851265} {\emph
  {\bibinfo {booktitle} {Studies in numerical analysis}}},\ \bibinfo {editor}
  {edited by\ \bibinfo {editor} {\bibfnamefont {G.~H.}\ \bibnamefont {Golub}}}\
  (\bibinfo  {publisher} {Mathematical Association of America},\ \bibinfo
  {year} {1984})\ p.~\bibinfo {pages} {3}\BibitemShut {NoStop}%
\bibitem [{\citenamefont {Mason}\ and\ \citenamefont
  {Handscomb}(2002)}]{Mason:2002}%
  \BibitemOpen
  \bibfield  {author} {\bibinfo {author} {\bibfnamefont {J.~C.}\ \bibnamefont
  {Mason}}\ and\ \bibinfo {author} {\bibfnamefont {D.~C.}\ \bibnamefont
  {Handscomb}},\ }\href
  {http://www.amazon.com/exec/obidos/redirect?tag=citeulike07-20\&path=ASIN/08%
49303559} {\emph {\bibinfo {title} {Chebyshev polynomials}}},\ \bibinfo
  {edition} {1st}\ ed.\ (\bibinfo  {publisher} {Chapman \& Hall/CRC},\ \bibinfo
  {year} {2002})\BibitemShut {NoStop}%
\bibitem [{\citenamefont {Green}(2008)}]{Green:2008}%
  \BibitemOpen
  \bibfield  {author} {\bibinfo {author} {\bibfnamefont {A.~M.}\ \bibnamefont
  {Green}},\ }\href {\doibase 10.1088/1475-7516/2008/07/005} {\bibfield
  {journal} {\bibinfo  {journal} {J. Cosmol. Astropart. Phys.}\ }\textbf
  {\bibinfo {volume} {06}},\ \bibinfo {pages} {005} (\bibinfo {year} {2008})},\
  \Eprint {http://arxiv.org/abs/0805.1704} {arXiv:0805.1704} \BibitemShut
  {NoStop}%
\bibitem [{\citenamefont {Strege}\ \emph {et~al.}(2012)\citenamefont {Strege},
  \citenamefont {Trotta}, \citenamefont {Bertone}, \citenamefont {Peter},\ and\
  \citenamefont {Scott}}]{Strege:2012}%
  \BibitemOpen
  \bibfield  {author} {\bibinfo {author} {\bibfnamefont {C.}~\bibnamefont
  {Strege}}, \bibinfo {author} {\bibfnamefont {R.}~\bibnamefont {Trotta}},
  \bibinfo {author} {\bibfnamefont {G.}~\bibnamefont {Bertone}}, \bibinfo
  {author} {\bibfnamefont {A.~H.~G.}\ \bibnamefont {Peter}}, \ and\ \bibinfo
  {author} {\bibfnamefont {P.}~\bibnamefont {Scott}},\ }\href
  {http://dx.doi.org/10.1103/PhysRevD.86.023507} {\bibfield  {journal}
  {\bibinfo  {journal} {Phys. Rev. D}\ }\textbf {\bibinfo {volume} {86}},\
  \bibinfo {pages} {023507} (\bibinfo {year} {2012})},\ \Eprint
  {http://arxiv.org/abs/1201.3631} {arXiv:1201.3631} \BibitemShut {NoStop}%
\end{thebibliography}%

\end{document}